\documentclass[useAMS,usenatbib]{mn2e}
\usepackage{amsmath,amssymb}
\usepackage{graphicx, epsfig, color}
\usepackage{natbib}
\usepackage{psfrag}
\usepackage{rotating}
\usepackage[abs]{overpic}
\title[Simulations of shocks encountering clumpy regions]{Numerical simulations of shocks encountering clumpy regions}

\author[R.\Aluzas ~et al.]
{R. \Aluzas$^{1}$\thanks{E-mail: js07ra@leeds.ac.uk}, J.~M.~Pittard$^{1}$, T.~W.~Hartquist$^{1}$, S.~A.~E.~G.~Falle$^{2}$, R.~Langton$^{1}$\\
$^{1}$School of Physics and Astronomy, University of
       Leeds, Woodhouse Lane, Leeds LS2 9JT, UK\\
$^{2}$Department of Applied Mathematics, University of 
	Leeds, Woodhouse Lane, Leeds LS2 9JT, UK
}
\date{\today}
\bibpunct{(}{)}{;}{a}{}{,}
\newcommand {\apgt} {\ {\raise-.5ex\hbox{$\buildrel>\over\sim$}}\ }
\newcommand {\aplt} {\ {\raise-.5ex\hbox{$\buildrel<\over\sim$}}\ } 

\def\eg{{\rm e.g.}}

\def\cf{{\rm cf.}}

\def\spose#1{\hbox to 0pt{#1\hss}}
\def\ltsimm{\mathrel{\spose{\lower 3pt\hbox{$\sim$}}
        \raise 2.0pt\hbox{$<$}}}
\def\gtsimm{\mathrel{\spose{\lower 3pt\hbox{$\sim$}}
        \raise 2.0pt\hbox{$>$}}}

\def\cm{{\rm\thinspace cm}}

\def\s{{\rm\thinspace s}}

\def\g{{\rm\thinspace g}}

\def\erg{{\rm\thinspace erg}}

\def\Hz{{\rm\thinspace Hz}}

\def\ster{{\rm\thinspace ster}}
\def\ergps{\hbox{${\rm\erg\s^{-1}\,}$}}

\def\pcm{\hbox{${\rm\cm^{-1}\,}$}}
\def\pcm2{\hbox{${\rm\cm^{-2}\,}$}}
\def\pcm3{\hbox{${\rm\cm^{-3}\,}$}}
\def\ergpscm3Hz{\hbox{${\rm\ergps\cm^{-3}\Hz^{-1}\,}$}}
\def\ergpscm3Hzster{\hbox{${\rm\ergps\cm^{-3}\Hz^{-1}\ster^{-1}\,}$}}
\def\gpcm3{\hbox{${\rm\g\cm^{-3}\,}$}}
\def\ergpcm2{\hbox{${\rm\erg\cm^{-2}\,}$}}
\def\ergpcm3{\hbox{${\rm\erg\cm^{-3}\,}$}}
\def\phpscm2{\hbox{${\rm photons\s^{-1}\cm^{-2}\,}$}}

\def\Aluzas{Al\={u}zas}

\begin{document}

\date{Accepted ... Received ...; in original form ...}


\maketitle


\begin{abstract}
  We present numerical simulations of the adiabatic interaction of a
  shock with a clumpy region containing many individual clouds. Our
  work incorporates a sub-grid turbulence model which for the first
  time makes this investigation feasible.  We vary the Mach number of
  the shock, the density contrast of the clouds, and the ratio of
  total cloud mass to inter-cloud mass within the clumpy region. Cloud
  material becomes incorporated into the flow. This ``mass-loading''
  reduces the Mach number of the shock, and leads to the formation of
  a dense shell. In cases in which the mass-loading is sufficient the
  flow slows enough that the shock degenerates into a wave.  The
  interaction evolves through up to four stages: initially the shock
  decelerates; then its speed is nearly constant; next the shock
  accelerates as it leaves the clumpy region; finally it moves at a
  constant speed close to its initial speed.  Turbulence is generated
  in the post-shock flow as the shock sweeps through the clumpy
  region. Clouds exposed to turbulence can be destroyed more rapidly
  than a similar cloud in an ``isolated'' environment. The lifetime of
  a downstream cloud decreases with increasing cloud-to-intercloud
  mass ratio. We briefly discuss the significance of these results for
  starburst superwinds and galaxy evolution.
\end{abstract}


\section{Introduction}
The energy input from high-mass stars in regions of vigorous
star formation creates bubbles of hot plasma
which may overlap to create superbubbles. Such structures are large enough that
they can burst out of their host galaxies, and vent mass and energy
into the intergalactic medium. However, the properties of such flows
may be dominated by their interaction with components with small
volume filling fractions. A complete understanding of the gas dynamics
of starburst regions, therefore, requires knowledge of how hot
plasma interacts with the cold, dense molecular material
present in the interstellar medium and in the superwind. 

Material in cold, dense clouds may be incorporated into
the hot phase via a variety of processes, including hydrodynamic
ablation and thermal or photoionization induced evaporation. This
``mass-loading'' is a key ingredient in models of galaxy formation and
evolution \citep[e.g.,][]{2010MNRAS.409.1541S}, but currently is not calculated
self-consistently in these models. The mass-loss
rates of clouds are also an important input parameter in models of the
ISM \citep{1977ApJ...218..148M}. Alternatively, clouds may be compressed
by the high pressure of the hot phase and collapse to form stars.

The response of a cloud to a given set of conditions remains to be
fully elucidated.  Although a large literature on interactions
involving single clouds exists
\citep[e.g.][]{1994ApJ...420..213K,2008ApJ...678..274O,2009MNRAS.394.1351P,2010MNRAS.405..821P},
there are only a handful of studies on the interaction of a flow
overrunning multiple clouds
\citep{1996ApJ...468L..59J,1997ApJ...491L..73S,2002ApJ...576..832P,2005MNRAS.361.1077P}. 
Thus, there remains a
clear need for further studies to examine the collective
effects of clouds on a flow, and to determine whether there are
differences between the behaviours of upstream and downstream clouds.

This is a first step in a long term study of the global effects of
clouds on an overruning flow. To simplify the problem and reduce the number
of free parameters we have assumed adiabatic behaviour, 
with a ratio of specific heats $\gamma= 5/3$. It alows us to follow the 
generic behaviour of a flow interacting with clouds. In future
we will perform calculations which include heating and radiative losses.
We investigate how cloud destruction affects
the density, speed and Mach number of the flow. We also determine whether
the position of a cloud within the clumpy region significantly
affects its evolution. In Section~\ref{sec:met} we describe the
numerical code and the initial conditions and a convergence
study which informs the subsequent work in this paper. Our results are
presented in Section~\ref{sec:res}, and conclusions and motivation for further
work are addressed in Section~\ref{sec:con}.

\section{Method}
\label{sec:met}
\subsection{Numerical Setup}

The computations were performed with the \emph{mg} hydrodynamic code.
This implements adaptive mesh refinement (AMR), a Godunov solver and
piecewise linear cell interpolation. The code solves numerically the
Euler equations of inviscid flow. These equations are supplemented by
a sub-grid turbulent viscosity model, which introduces two additional
variables; $k$, the turbulent kinetic energy, and $\epsilon$, its
dissipation rate. The fluxes of $k$ and $\epsilon$ across the
cell boundaries are also returned by the Riemann solver. The values of $k$ and $\epsilon$
in each cell are updated accordingly, including changes due to source terms.
The full set of equations solved can be found in \citet{2009MNRAS.394.1351P}. 
The energy source term in eq. 7 of  \citet{2009MNRAS.394.1351P}, $S_E$, is $-S_k$, defined in eq. 13.
$T$ is temperature and $\tau$ is the turbulent stress tensor.  More details about
the implementation of the $k$-$\epsilon$ model can be found in \cite{1994MNRAS.269..607F}.

The adopted grids are cartesian, and either 2-D, for clouds that are initially
infinite cylinders, or 3-D, for clouds that are initially spheres. We first performed a
resolution study for the interaction of a flow with a single
cloud. Constant inflow was imposed from the
negative $x$-direction together with zero-gradient conditions on the other
boundaries. The simulations were terminated before any of the cloud material
reached the edges of the grid. In the multi-cloud simulations we imposed
periodic boundary conditions in the $y$-direction, with the same inflow and free outflow
in the $x$-direction as in the single cloud case.

Two grids ($G^0$ and $G^1$) cover the entire domain. Finer grids were
added where they were needed and removed where they were not, based on
the refinement criteria controlled by differences in the solutions on
the coarser grids. Each refinement level increased the resolution in
all directions by a factor of 2. The time-step on grid $G^n$ was
$\Delta t_0 / 2n$ where $\Delta t_0$ was the time-step on $G^0$. 
The presence of the $k$-$\epsilon$ subgrid model adds a viscous stability 
condition. The time step is then the smaller of the viscous and hyperbolic
time steps. An explicit method is used to advance the solution forwards in time.
In single cloud resolution tests, $4$ to $8$ levels of refinement were
used with the coarsest $G^0$ grid covering the cloud with $0.5-2$
cells per cloud radius, in order to achieve an effective resolution of
up to 256 cells per cloud radius. The effective resolution is given by
the resolution in the finest grid and is quoted as $R_{cr}$, where
\emph{cr} is the number of cells per cloud radius in the finest
grid. For example, $R_{256}$ indicates an effective resolution of
$256$ cells per cloud radius. Four levels of refinement were used in
the multiple cloud simulations to achieve a resolution of $R_{8}$. All
length scales are measured in units of a cloud radius.

The clouds and ambient medium are initially at uniform pressure. Cloud edges 
have a density profile from \citet{2009MNRAS.394.1351P}, 
with $p_1 = 10$. In the single cloud simulations the cloud was centered on the grid
origin, and a planar shock front was imposed at $x=-3$. Time zero is defined
to be the time when the shock first encounters the cloud.
Each simulation is described by the Mach number of the shock, $M$, and the cloud density
contrast, $\chi$. The time is measured in units of the cloud
crushing timescale, $t_{\rm cc} = \chi^{1/2} r_{\rm cl}/v_{\rm b}$, where
$r_{\rm cl}$ is the initial cloud radius and $v_{\rm b}$ is the shock velocity in
the ambient medium \citep{1994ApJ...420..213K}. Kelvin-Helmholtz (KH)
and Rayleigh-Taylor (RT) instabilities destroy the cloud in $\approx
10\,t_{\rm cc}$ for strong-shock interactions, with weaker shocks
taking longer \citep[see][]{2010MNRAS.405..821P}.

In the multiple cloud simulations the clouds are initially randomly
distributed within a rectangular region which we refer to as the
``clumpy region''. Each cloud has an identical radius and density
contrast, and because of the 2D geometry each cloud represents an
infinite cylinder. The size of this region is $400\,r_{\rm (cl)}
\times 400\,r_{\rm (cl)}$, unless otherwise noted. A key parameter in
the multi-cloud simulations is the ratio of total cloud mass to
inter-cloud mass in the clumpy region, $MR$. In this work, mass ratios
of $MR=0.25 -4$ are investigated. Time zero occurs when the shock is
just outside the clumpy region, but is often shifted in the subsequent
analysis to coincide with the shock entering the region of interest.
The clumpy region is divided into 4 equally sized blocks 
(shown in Fig.~\ref{fig:m3w6_early}) with block 1
being furthest upstream and the first to be hit by the shock. Various
properties of the total cloud material within a given block and select
individual clouds within each block were monitored through the use of
advected scalars which ``colour'' the flow.  In pasticular, the
velocity and mass of cloud material in individual blocks and clouds
are studied.  An algorithm which searches for the cells which are
furthest downstream and contain a pressure higher than the ambient
pressure is used to track the position of the shock as a function of
time.

\begin{figure*}
  \begin{center}
    \begin{tabular}{ccc}
\psfig{figure=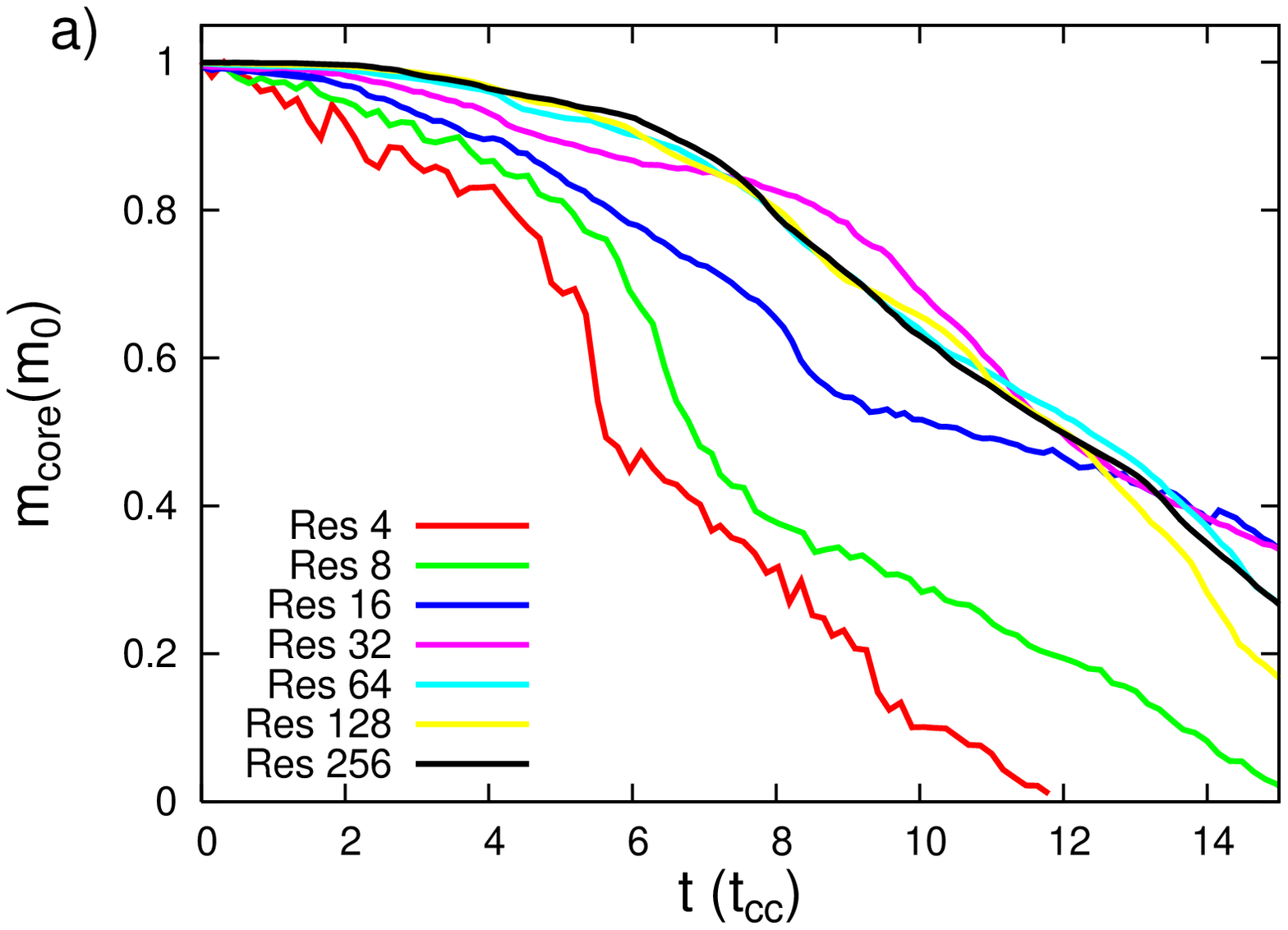, angle=-90, width=5.5cm} &
\psfig{figure=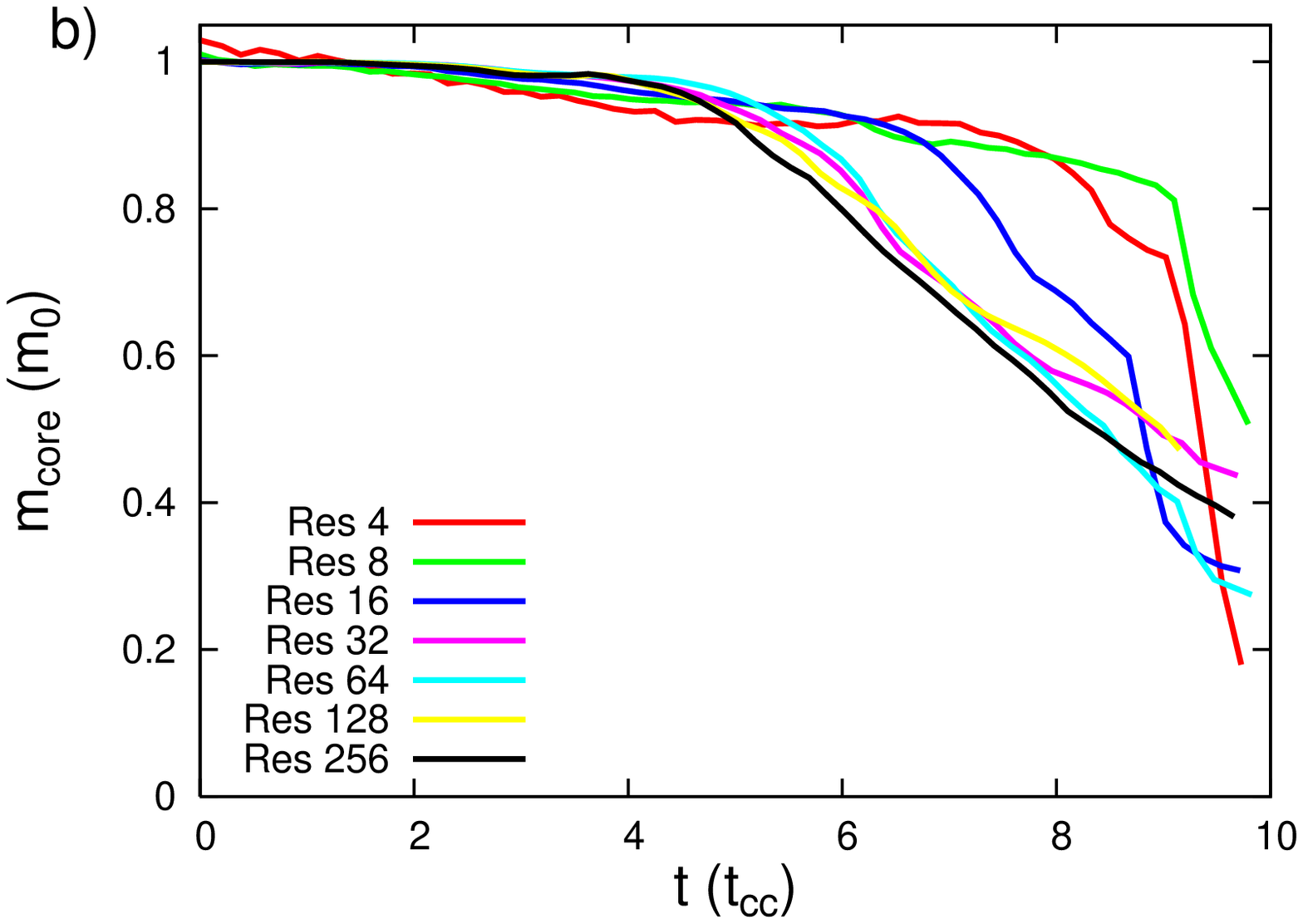, angle=-90, width=5.5cm} &
\psfig{figure=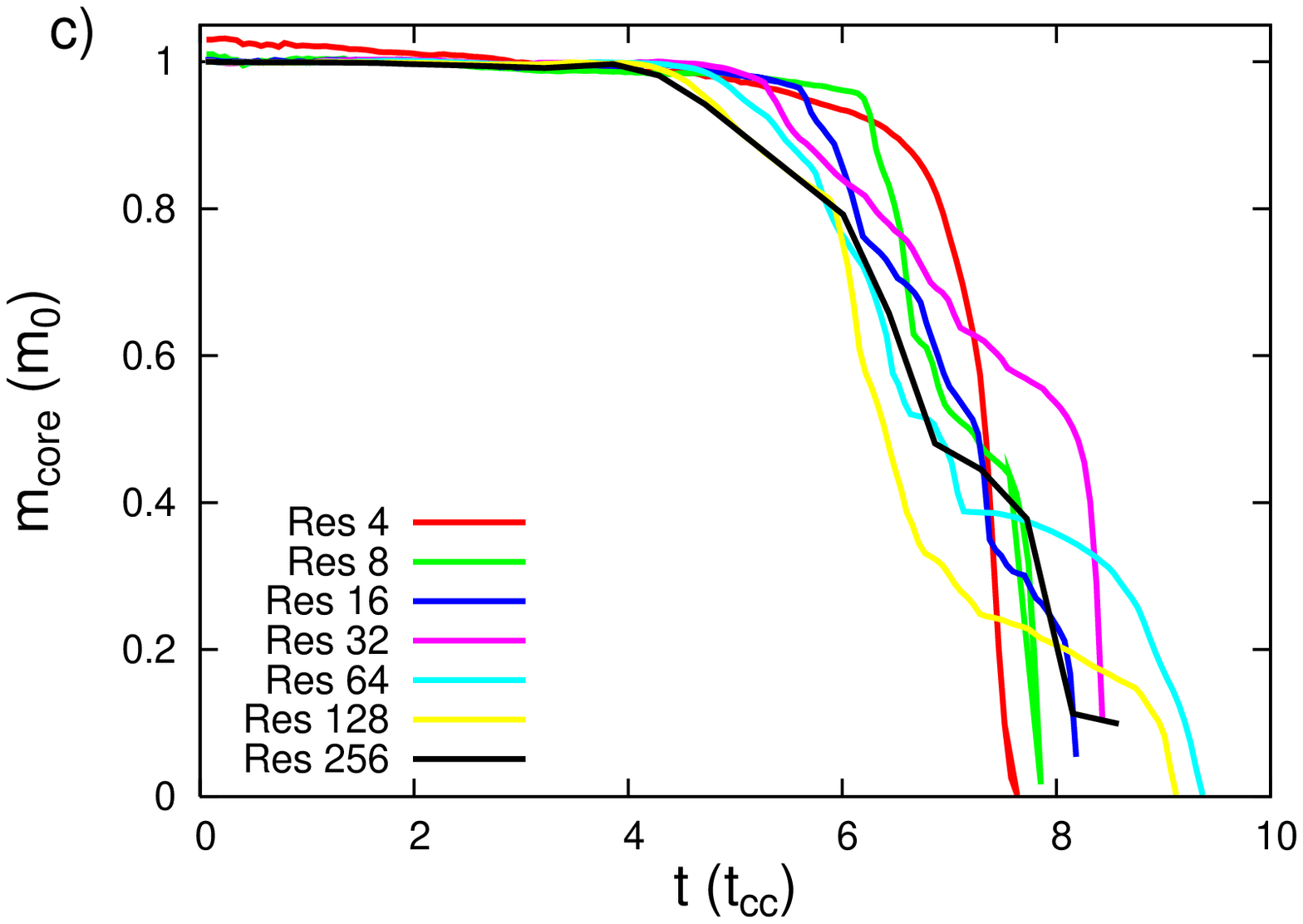, angle=-90, width=5.5cm} \\
\psfig{figure=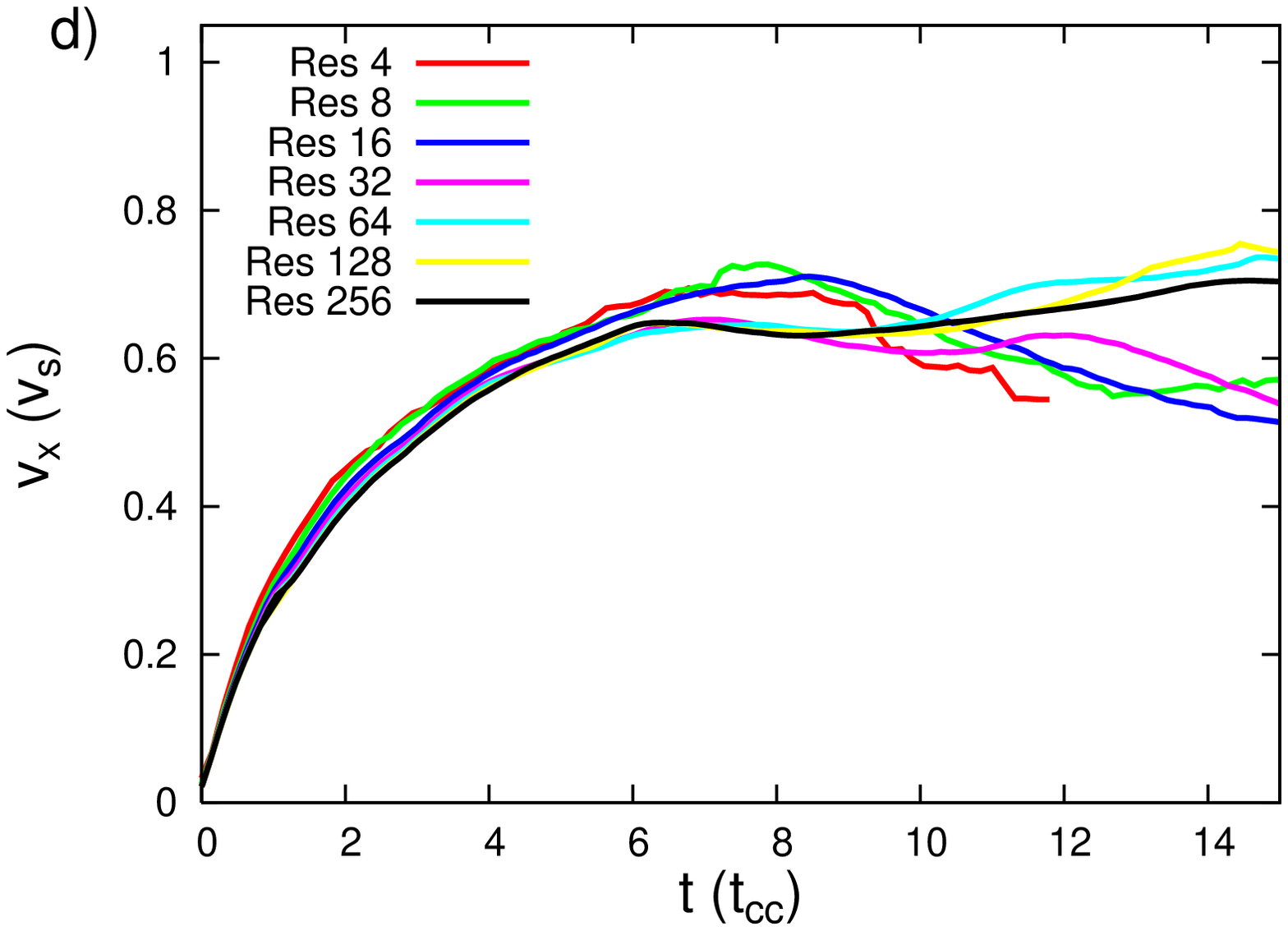, angle=-90, width=5.5cm} &
\psfig{figure=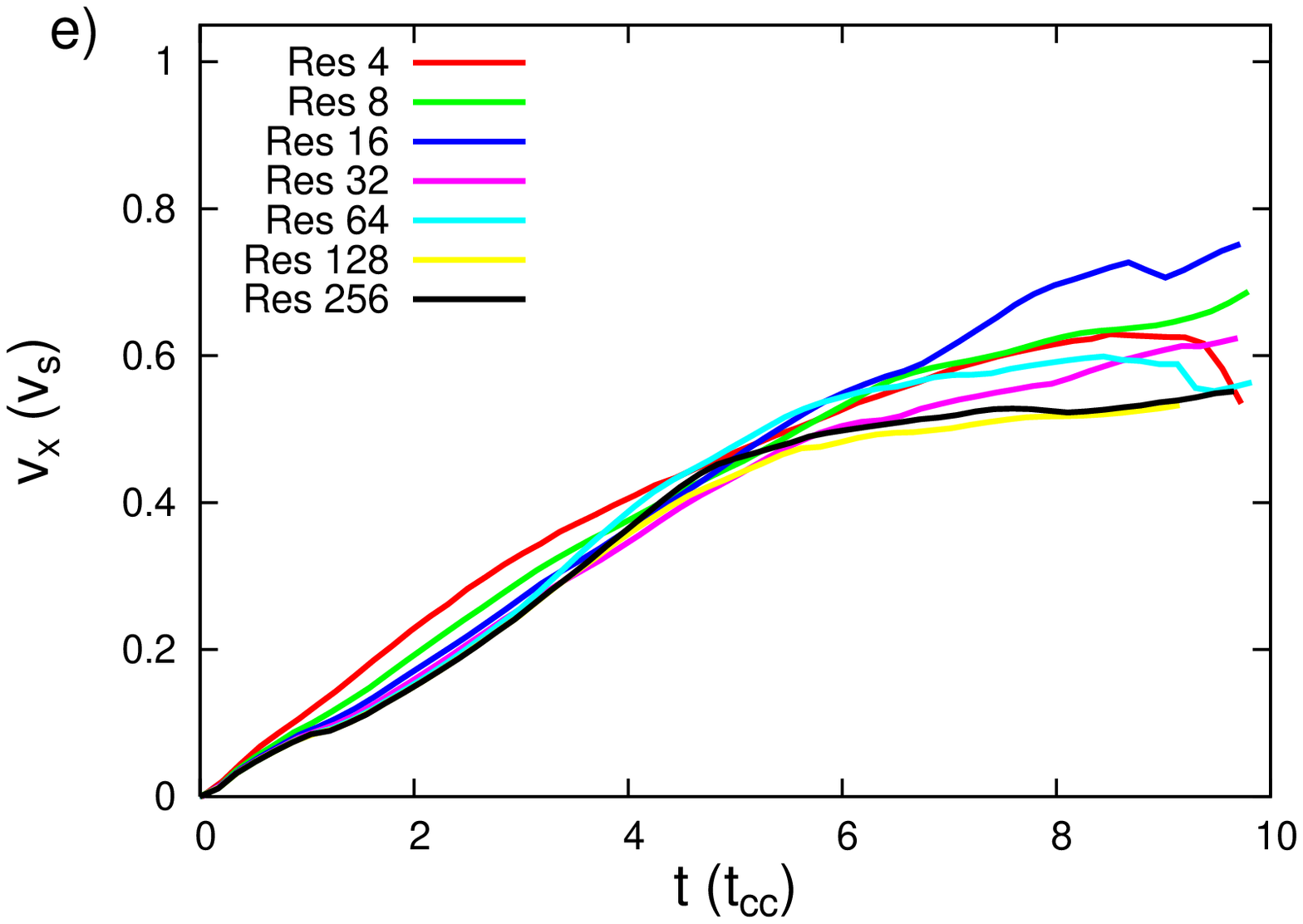, angle=-90, width=5.5cm} &
\psfig{figure=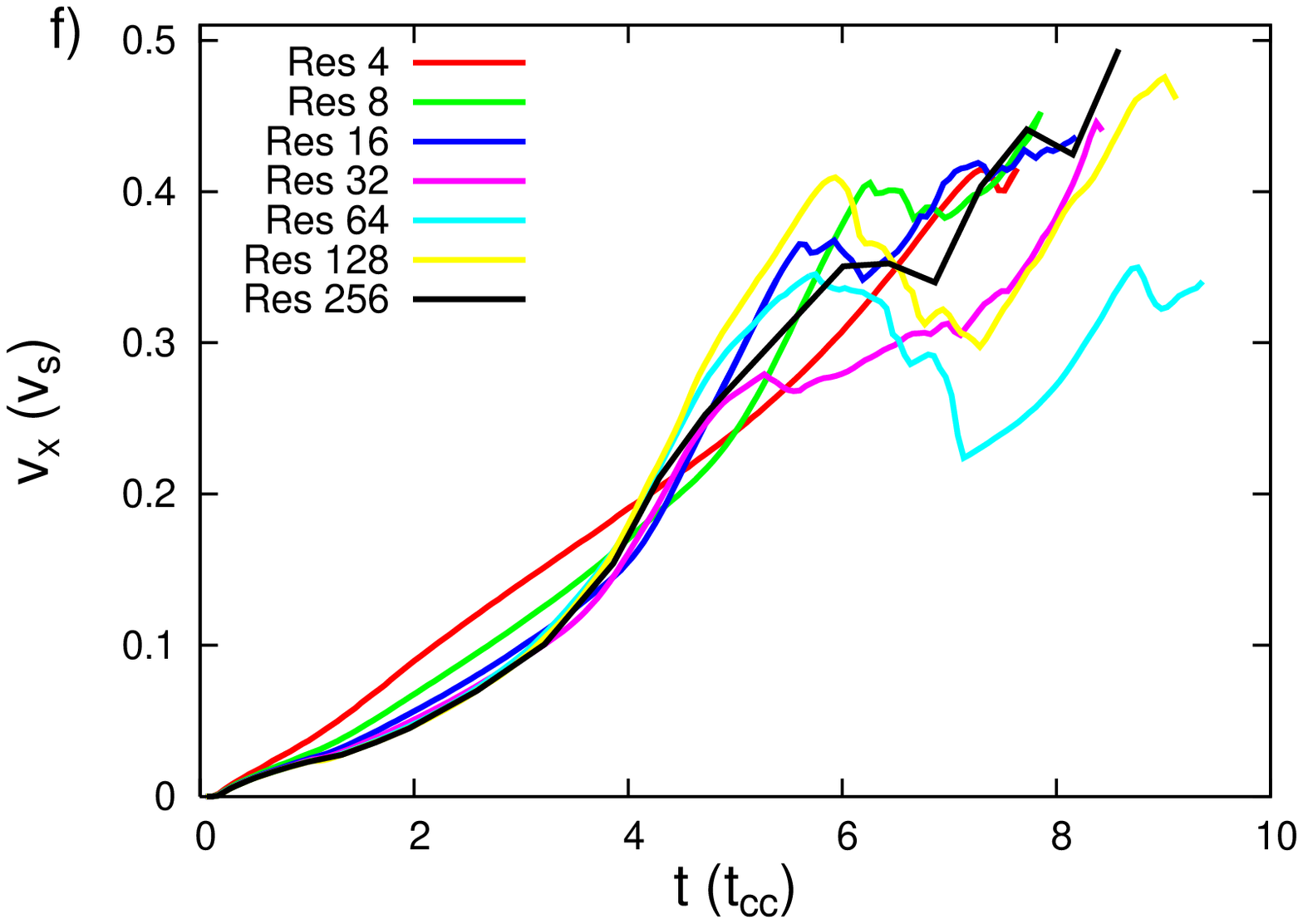, angle=-90, width=5.5cm} \\
    \end{tabular}
    \caption{Convergence tests for 2D simulations of a
      Mach 5 shock hitting a cylindrical cloud with density contrast $\chi=10$ (left
      column), $\chi=10^2$ 
      (center column), and $\chi=10^3$ (right column). 
      The time evolution of the core mass (top row) and the mean
      cloud velocity (bottom row) are shown.  }
    \label{fig:res5}
  \end{center}
\end{figure*}

\subsection{Convergence Studies}
Previous resolution tests of numerical shock-cloud interactions
\citep[\eg,][]{2006ApJS..164..477N}) have revealed that adiabatic,
hydrodynamic simulations need $\sim100$ cells per cloud radius
($R_{100}$) for a converged result.  Simulations including more
complex physics, especially ones with strong cooling, have been found
to need higher resolutions \citep{2010ApJ...722..412Y}.  In contrast,
\citet{2009MNRAS.394.1351P} found that adiabatic simulations with a
$k$-$\epsilon$ subgrid turbulence model show better convergence, and
(most importantly for this work) converge at lower resolutions.

Currently, multiple cloud simulations cannot be performed at a resolution 
as high as $R_{100}$ and 3D simulations are also much more
expensive computationally. The first (and to our knowledge, only)
resolution tests of a multiple cloud interaction were performed by
\cite{2004ApJ...604..213P}, who showed how resolutions of $R_{16}$,
$R_{32}$ and $R_{64}$ affect the shock position at a given time. They
concluded that the differences between their results were smaller than
the sensitivity of their proposed experimental design so no further
analysis was done.  All the more extensive resolution tests in the
current astrophysical literature concern 2D simulations of an
individual cloud.  We have therefore performed various resolution
tests including 2D single cloud simulations with and without a
$k$-$\epsilon$ model for different values of $\chi$ and $M$. 
A 3D single cloud simulation with $M=5$ and $\chi=10^2$
was investigated with different resolutions up to $R_{64}$. Multiple
(13 and 48) cloud simulations with $M=3$, $\chi=10^2$ and $MR=$ 1 and
4 were also performed at resolutions $R_4$, $R_8$, $R_{16}$ and
$R_{32}$.

We focus on two measures which are affected by the mixing of cloud
material into the flow. These are the mean cloud velocity ($< v_{x}>$)
and the core mass of the cloud ($m_{\rm core}$). The latter is defined as
the sum total of cloud mass in grid cells where more than half the cell's
mass is cloud material. 

\subsubsection{Single cloud resolution tests}
Fig.~\ref{fig:res5} shows the evolution of $m_{\rm core}$ and $< v_{x}>$
for 2D simulations of cylindrical clouds hit by an
$M=5$ shock for a range of values of $\chi$.

\begin{figure}
\begin{center}
\begin{tabular}{c}
\psfig{figure=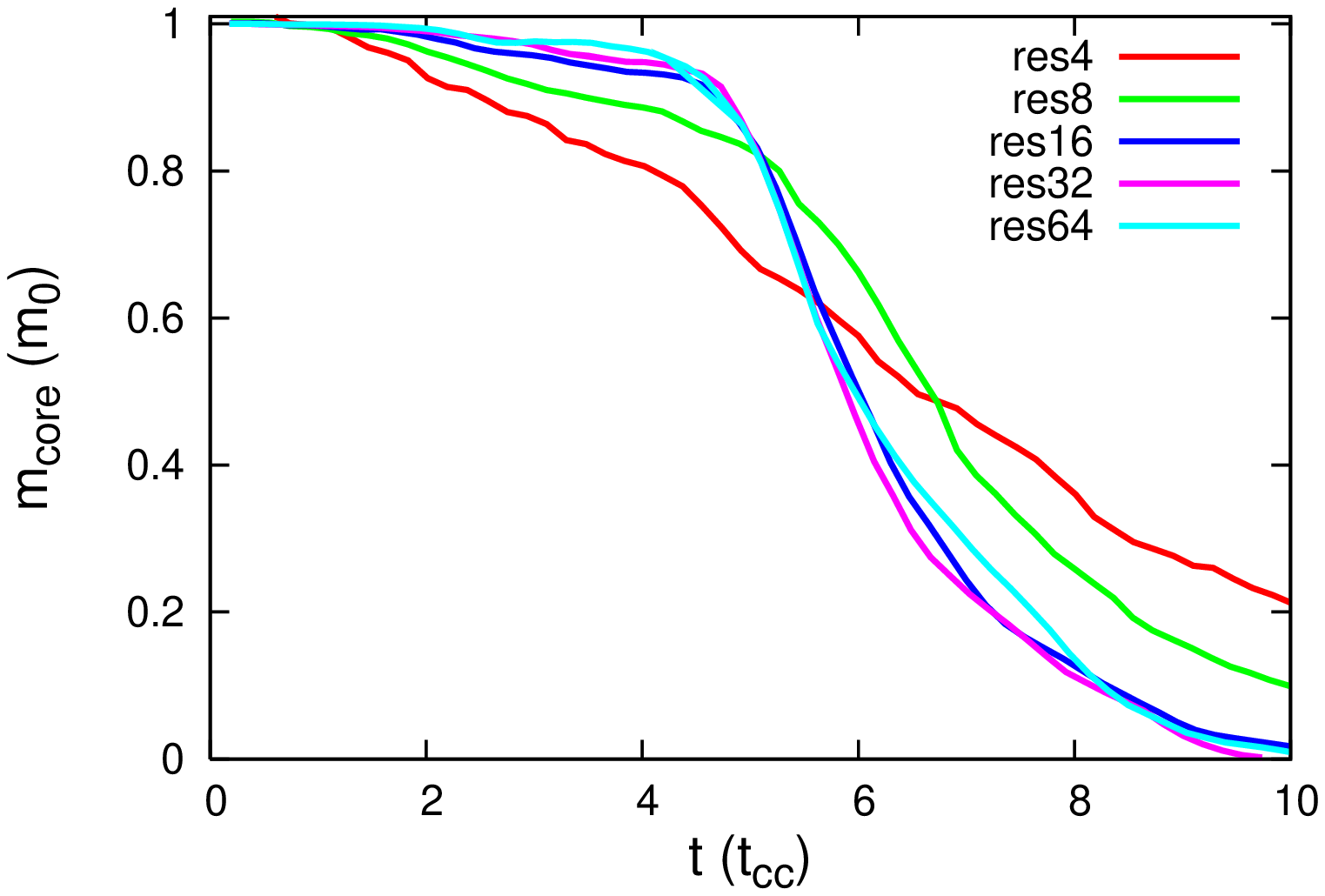, angle=-90, width=6.5cm} \\
\psfig{figure=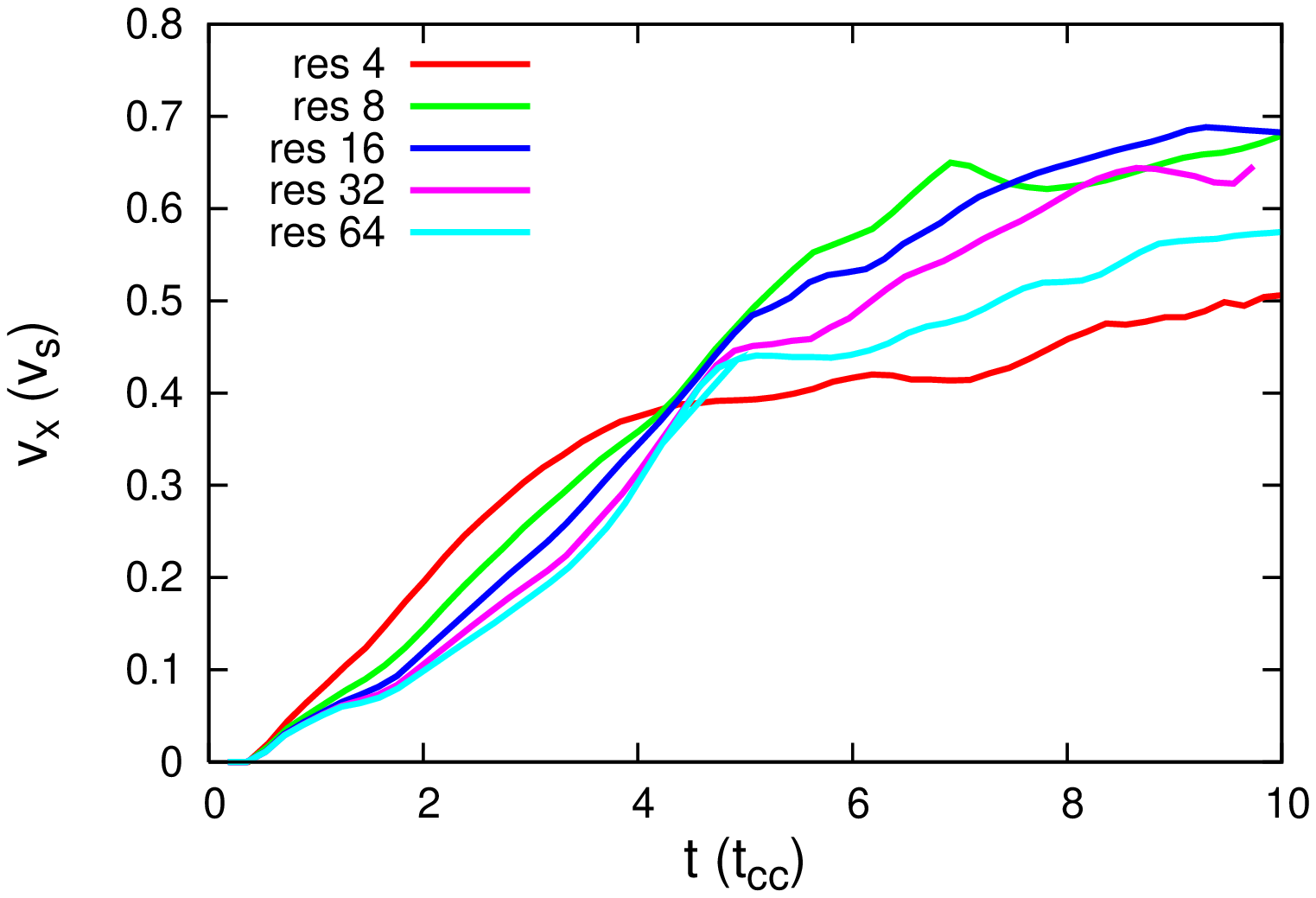, angle=-90, width=6.5cm}
\end{tabular}
\caption[]{Convergence tests for 3D simulations of a
      Mach 5 shock hitting a spherical cloud with density contrast $\chi=10^2$. 
      The time evolution of the core mass (top) and the mean
      cloud velocity (bottom) are shown. }
\label{fig:res_3D}
\end{center}
\end{figure}

\begin{figure}
\begin{center}
\begin{tabular}{c}
\psfig{figure=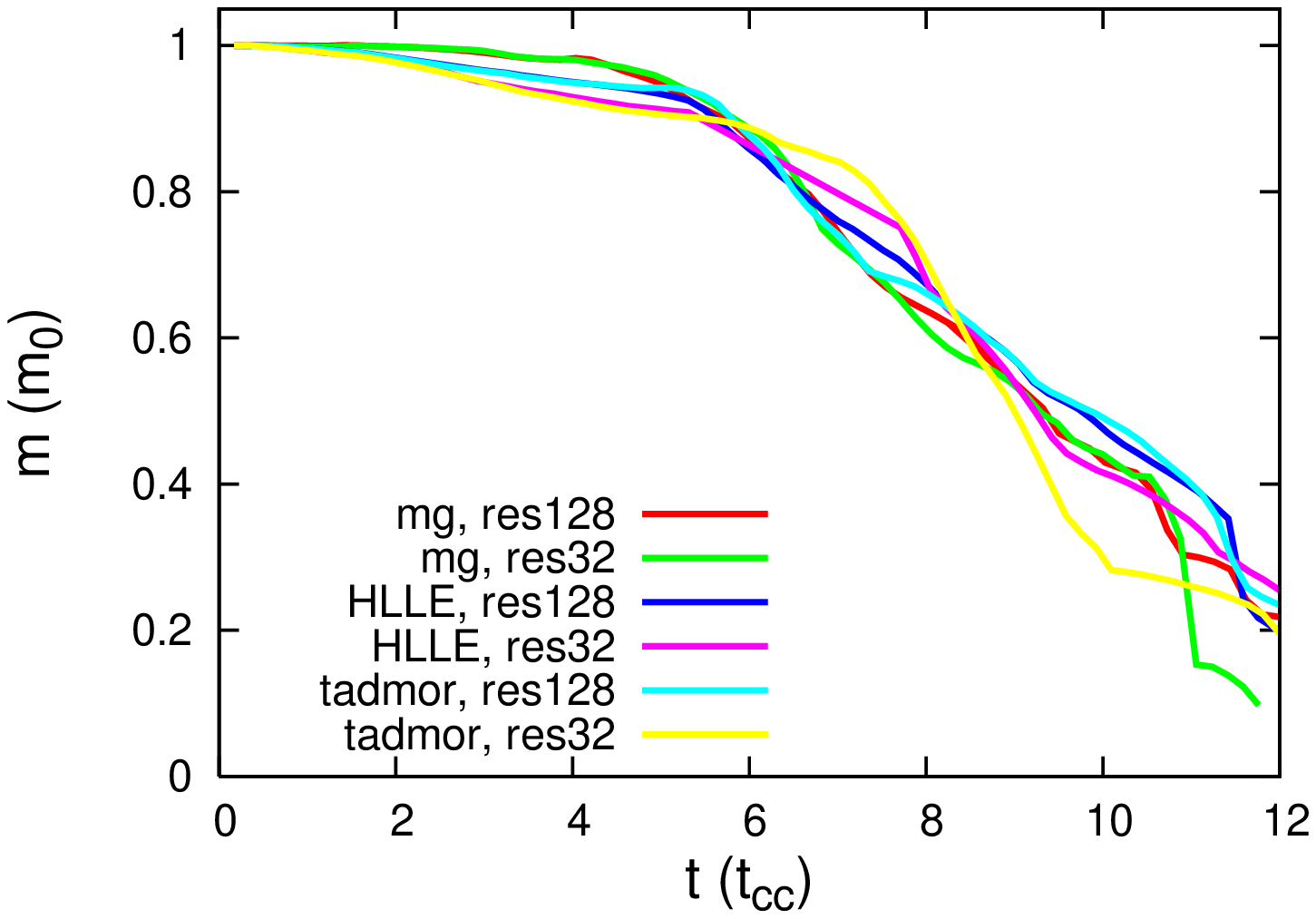, angle=-90, width=6.5cm} \\
\psfig{figure=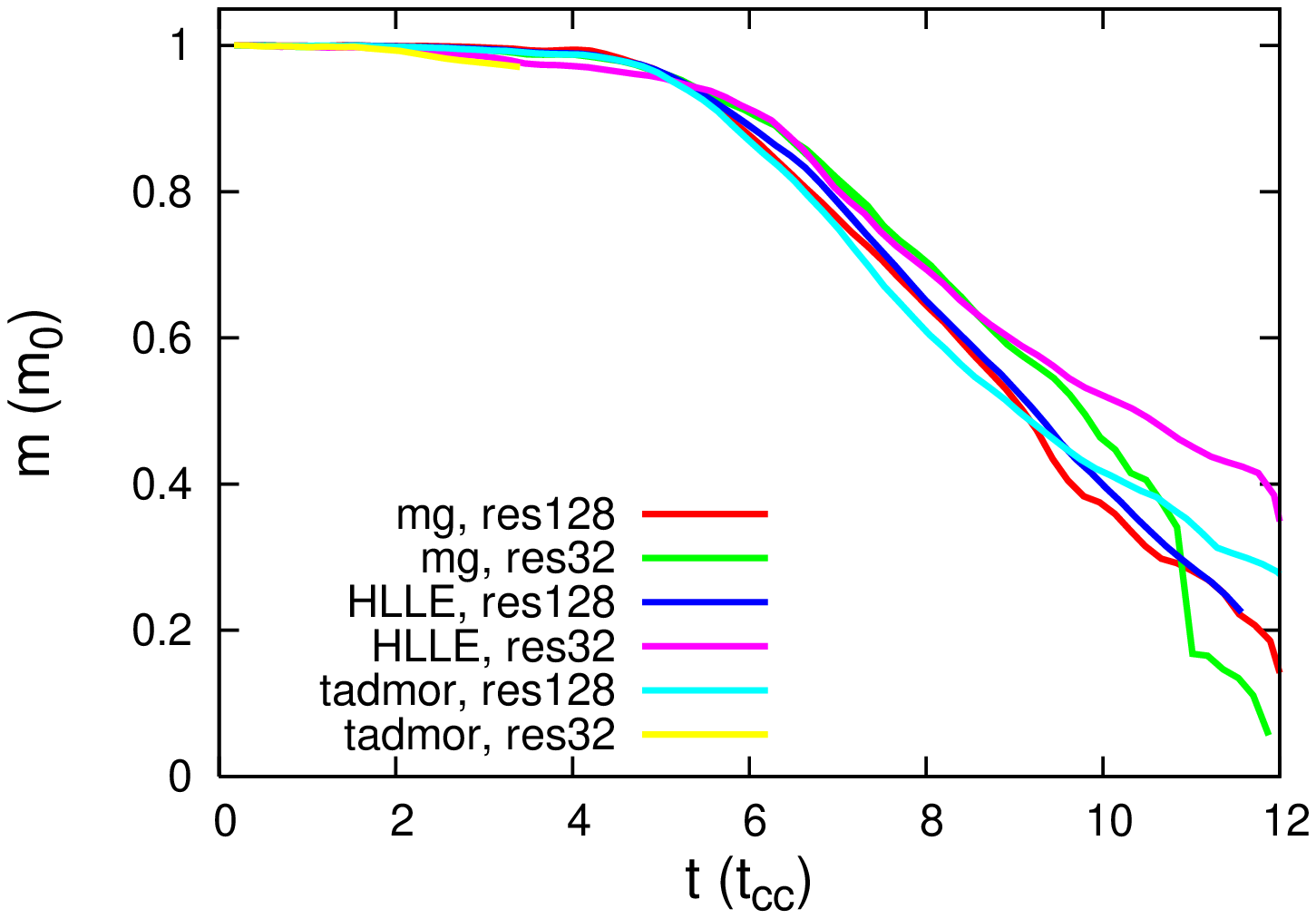, angle=-90, width=6.5cm}
\end{tabular}
\caption[]{Comparison of the core mass evolution when using the linear/exact 
Riemann solver in mg and other Riemann solvers. 
The $k$-$\epsilon$ subgrid model is enabled in the top panel and disabled in the bottom panel.
We experienced difficulties with one of the Tadmor calculations which stops at $t=3\,t_{\rm cc}$.}
\label{fig:riemcomp}
\end{center}
\end{figure}

In all cases the interaction leads to most of the
core mass being mixed with ambient material by $t \approx
10\,t_{\rm cc}$. The simulations at lower density contrasts are most
resolution dependent. When $\chi=10$ the cloud mixes faster at lower
resolutions. The opposite is true when $\chi=10^{2}$. For $\chi
\ltsimm 10^{2}$, simulations with at least 32 cells per cloud radius
are much better converged than simulations at lower resolution. For
the highest density contrast studied ($\chi=10^3$) convergence is obtained at
lower resolution.

We now examine the resolution dependence of a spherical cloud in a 3D simulation.
Fig.~\ref{fig:res_3D} shows the evolution of the core mass and mean
cloud velocity as a function of resolution for a cloud with
$\chi=10^{2}$ hit by a Mach 5 shock.
While the velocity diverges a bit after $5\,t_{\rm cc}$, 
the core mass evolution at $R_{16}$ 
exhibits the same features as at $R_{32}$ and $R_{64}$, whereas in 2D
a resolution of $R_{32}$ is required for a similar level of
convergence. Thus, it appears that the extra degrees of freedom for
instabilities in 3D versus 2D has the effect that convergence is
achieved at lower resolutions.

The geometry makes a significant difference 
to the cloud evolution. Panel b of Fig.~\ref{fig:res5}
shows that a cylindrical cloud has lost about half of its core mass
by $t \approx 8\,t_{\rm cc}$, whereas Fig.~\ref{fig:res_3D} shows
that for a spherical cloud this occurs by $t\approx 6\,t_{\rm cc}$.

If the time-averaged pressure returned by the linear solver differs by less than 10\% from the
pressures in the left and right states, the solution from the linear solver is used. Otherwise an exact solver
with the standard Secant method is used. We have compared the effects of different Riemann solvers on single
cloud simulations. In particular we have used HLLE \citep{1988SJNA...25..294E} and Tadmor \citep{1990JCoPh..87..408N} solvers. With a first 
order (spatial and temporal) update the scheme is very diffusive irrespective of which Riemann solver is used
 (although the \emph{mg} solver is least diffusive).  Fig.~\ref{fig:riemcomp} shows results with the standard
second order scheme. It reveals that the different solvers produce similar results at high resolutions, while
the $k$-$\epsilon$ model reduces the resolution dependance, especially at late times 
\citep[\cf \,][]{2009MNRAS.394.1351P}.

\begin{figure}
\begin{center}
\includegraphics[width=0.67\columnwidth]{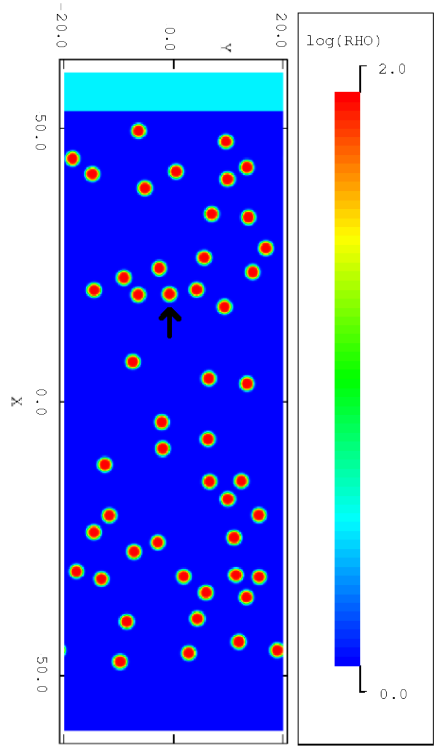}
\end{center}
\caption{The initial ($t=0$) cloud distribution for a simulation of a shock
  overrunning multiple cylindrical clouds. The shock Mach number $M=3$, the cloud density contrast $\chi =
  10^2$, and the ratio of cloud mass to intercloud mass in the clumpy
  region is $MR=4$. The arrow marks an individual cloud for which the
  properties were monitored in time - see Fig.~\ref{fig:res_mc1}.
  In this plot and in all other density snapshots the spatial scale is in the units of the cloud radius and the density is in the units of the ambient density.
   }
\label{fig:mc_res_t0}
\end{figure}

\begin{figure}
\begin{center}
\includegraphics[width=\columnwidth]{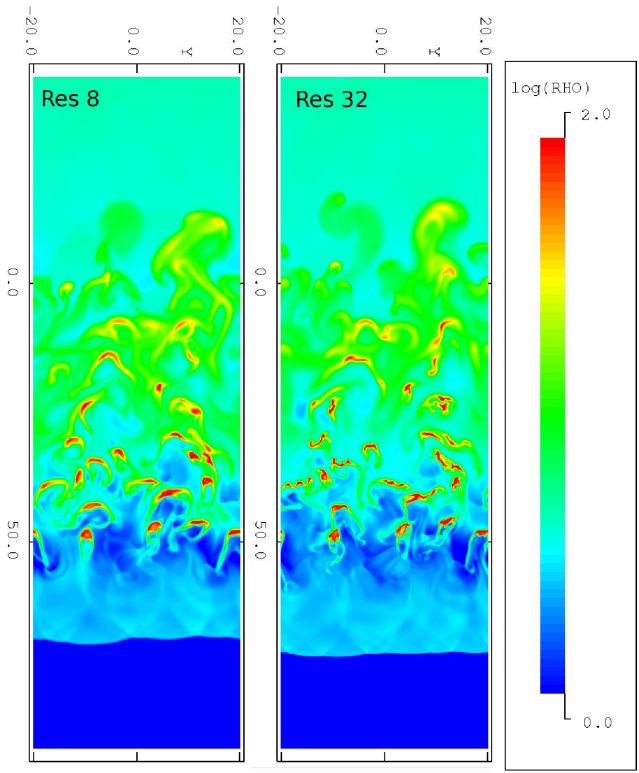}
\caption{Resolution test for a Mach 3 shock overrunning multiple
    cylindrical clouds with the initial setup shown in
  Fig.~\ref{fig:mc_res_t0}. A logarithmic density plot is shown at
  $t=14.2\,t_{\rm cc}$ for resolutions of $R_{8}$ (left) and $R_{32}$ (right).
 }
\label{fig:res_pic}
\end{center}
\end{figure}

\begin{figure}
  \begin{center}
    \begin{tabular}{c}
      \psfig{figure=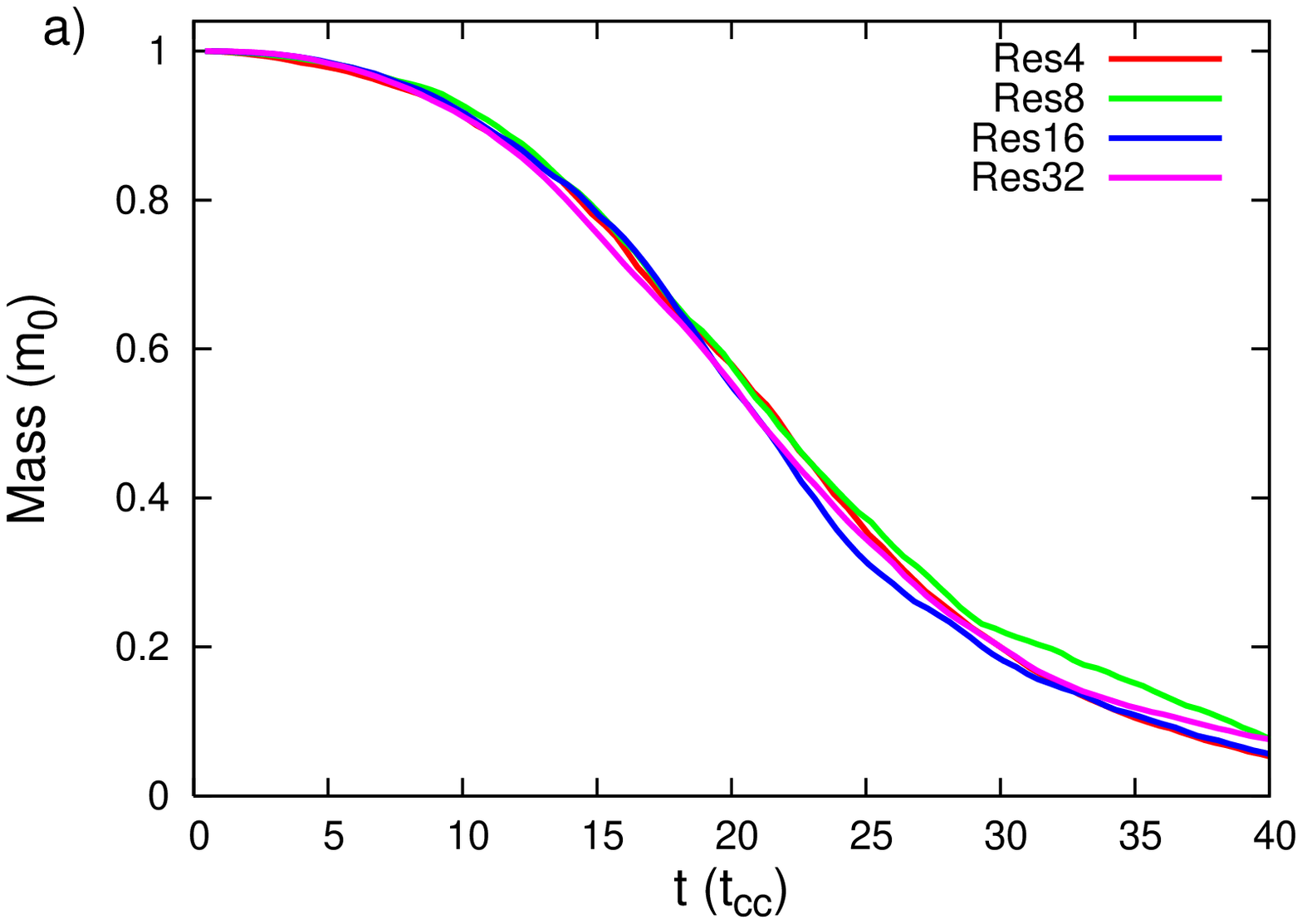, angle=-90, width=65mm} \\
      \psfig{figure=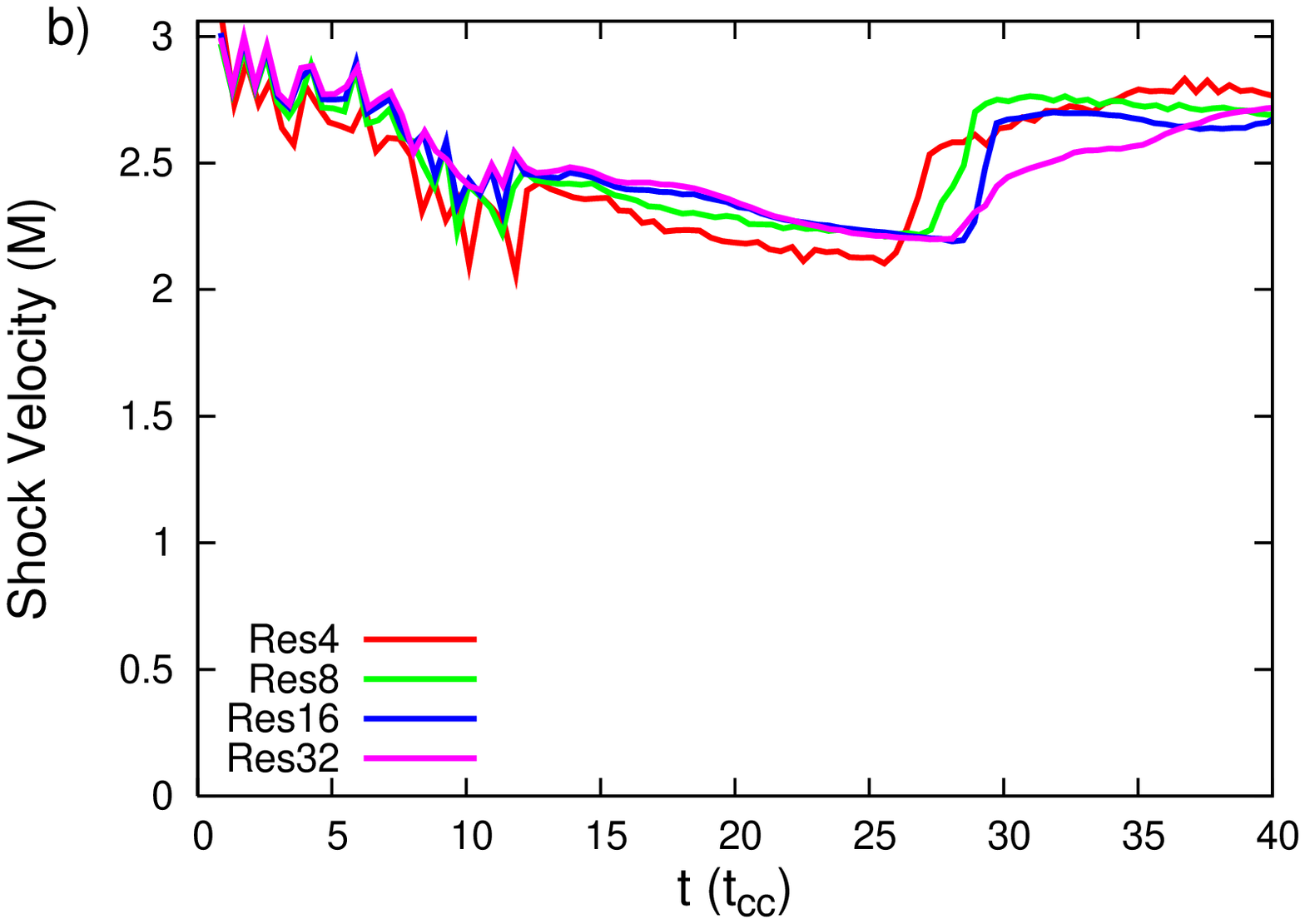, angle=-90, width=65mm} \\
      \psfig{figure=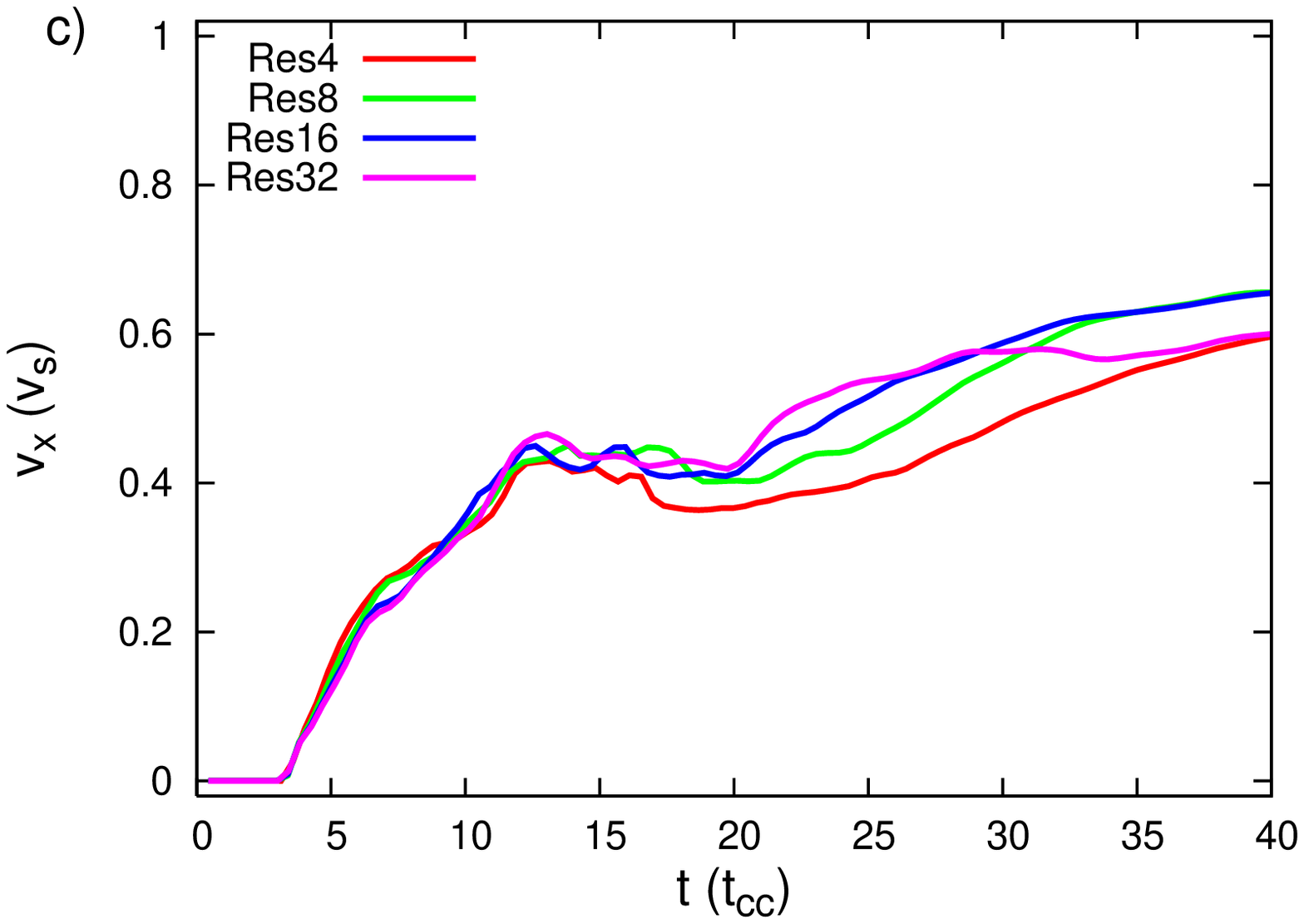, angle=-90, width=65mm} \\
    \end{tabular}
  \end{center}
  \caption{Resolution test for a Mach 3 shock overrunning multiple
  cylindrical clouds with $\chi=10^{2}$ and $MR = 4$ (see Fig.~\ref{fig:res_pic}).
  The time evolution of a) the total mass
  of all of the cloud cores, b) the mean shock velocity
  (normalized to the ambient intercloud sound speed)
  and c) the average core velocity of a single cloud, are shown. 
  }
    \label{fig:res_mc1}
\end{figure}

\subsubsection{Multiple clouds resolution test}
\label{sec:res_mc}
We have also performed a convergence study for 2D multi-cloud
simulations. In this study a Mach 3 shock overruns 48 identical
cylindrical clouds, each with a density
contrast to the intercloud medium of $\chi = 10^2$. The ratio of cloud
mass to intercloud
mass within the clumpy region is $MR=4$. Fig.~\ref{fig:mc_res_t0} shows the initial
setup and distribution of clouds. The clouds fill a region which is
$100\,r_{\rm cl}$ deep and $40\,r_{\rm cl}$ wide, with periodic
boundaries in $y$. Resolutions of $R_{4}$, $R_{8}$, $R_{16}$ and
$R_{32}$ were used. The latter required a grid with an effective
size of $12800\times 1280$. 

The shock exits the clumpy region at around $t=12.5\,t_{cc}$ in all of
the simulations. Density plots for times just after this moment are shown in
Fig.~\ref{fig:res_pic} for the $R_{8}$ and $R_{32}$ simulations.
Higher compressions and greater fragmentation are seen in the higher
resolution simulation.  However, while significant differences in the
behaviour of any individual cloud in the simulation can be identified,
Fig.~\ref{fig:res_mc1} shows that important \emph{global} parameters
in the multiple cloud simulation, such as the rate at which cloud mass
is mixed into the flow (shown as the time evolution of the total mass
of all of the cloud cores in Fig.~\ref{fig:res_mc1}a), the rate at
which momentum is transferred from the flow to the clouds (shown as
the velocity of a single cloud, Fig.~\ref{fig:res_mc1}c), and shock
velocity (Fig.~\ref{fig:res_mc1}b) are {\em not} very sensitive to the
resolution used.

Indeed, it is clear from Fig.~\ref{fig:res_mc1} that a resolution of
$R_{8}$ is sufficient in order to obtain an accurate
representation of the global effect of multiple clouds on a flow. In
previous multi-cloud simulations resolutions of $R_{32}$ and $R_{16}$
have been adopted (\cite{2002ApJ...576..832P} and
\cite{2004ApJ...604..213P}, respectively), but we show here that a
somewhat lower resolution can be safely used, at least when a sub-grid
turbulence model is employed.  Therefore, we perform all other
simulations in this work at a resolution of $R_8$.

\section{Results}
\label{sec:res}
In this section we show the results of a number of simulations with
different values of $M$, $\chi$, and $MR$.  Table \ref{table:sims} 
summarizes the simulations that were
performed.  We adopt a naming convention such that \emph{m3chi2MR1}
refers to a simulation with $M=3$, $\chi=10^2$ and $MR=1$.

\begin{table}
\caption{Summary of the multi-cloud simulations performed.} 
\centering 
\begin{tabular}{c c c c c} 
\hline\hline 
Simulation & $\chi$ & $MR$ & ncc\footnotemark[1] & Shock Mach number \\ [0.5ex] 
\hline 
chi2MR0.25 & $10^2$ & 0.25 & 128 & 3\\ 
chi2MR1 & $10^2$ & 1 & 505 & 1.5, 3, 10\\
chi2MR1\_double\footnotemark[2] & $10^2$ & 1 & 1110 & 3 \\
chi2MR4 & $10^2$ & 4 & 1959 & 1.5, 2, 3, 10 \\
chi3MR0.25 & $10^3$ & 0.25 & 13 & 3 \\
chi3MR1 & $10^3$ & 1 & 51 & 3 \\
chi3MR4 & $10^3$ & 4 & 203 & 3 \\ [1ex] 
\hline 
\end{tabular}
\label{table:sims} 

\end{table}

\footnotetext[1]{Number of clouds in the clumpy region.}
\footnotetext[2]{Simulation \emph{chi2MR1\_double} has the same cloud distribution as
\emph{chi2MR1} but it is repeated in the $x$-direction to obtain a
distribution with twice the depth along the direction 
of shock propagation.}

\subsection{Mach 3 shock interactions}
\begin{figure*}
\includegraphics[width=\textwidth]{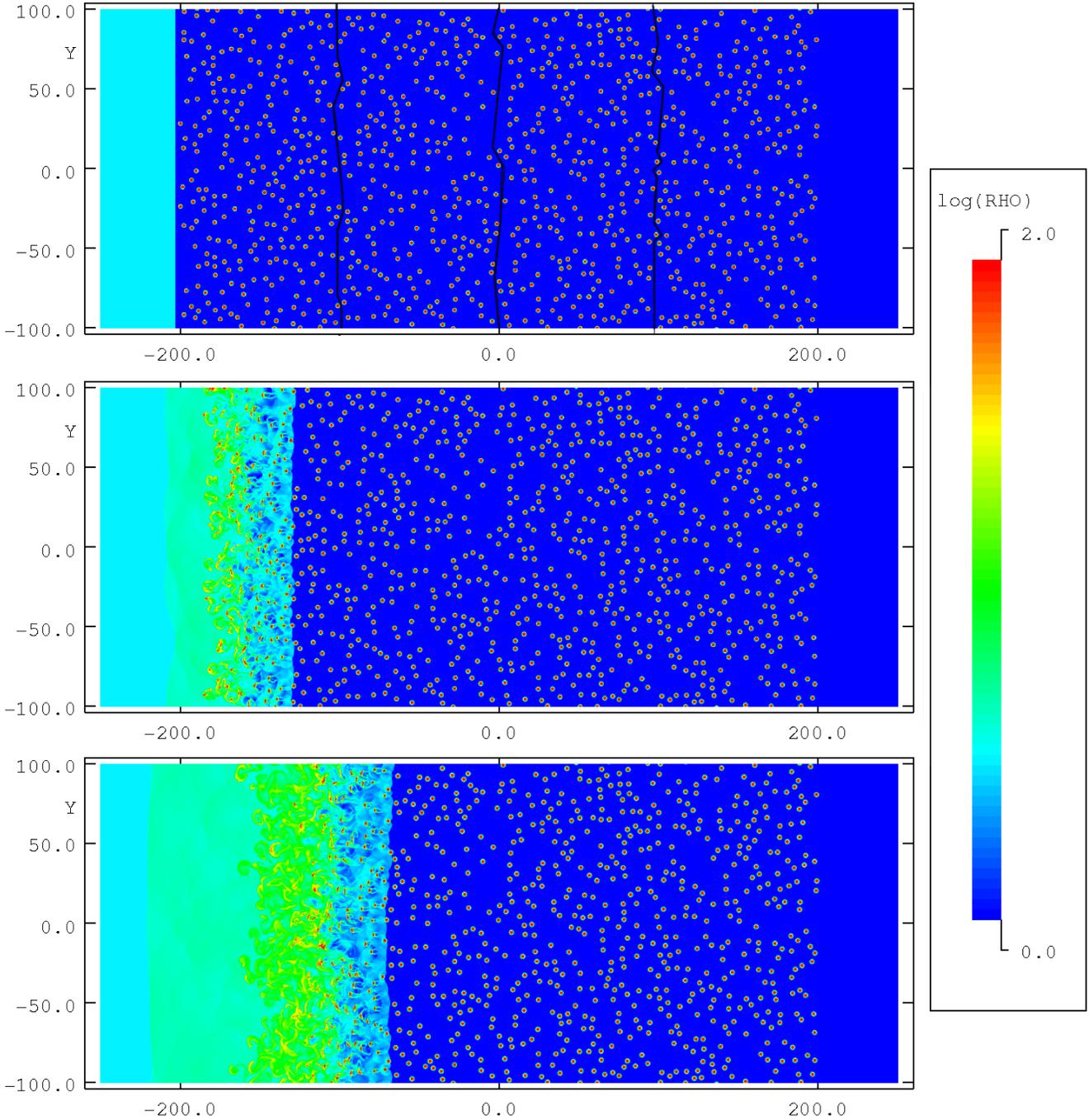}
\caption{The time evolution of the logarithmic density for model
  \emph{m3chi2MR4}, shown at $t=0$ (top), $t=8.1\,t_{\rm cc}$ (middle) and 
$t=16.2\,t_{\rm cc}$ (bottom). The top panel also shows how the
different cloud regions are defined.
 }
\label{fig:m3w6_early}
\end{figure*}

\begin{figure*}
\includegraphics[width=\textwidth]{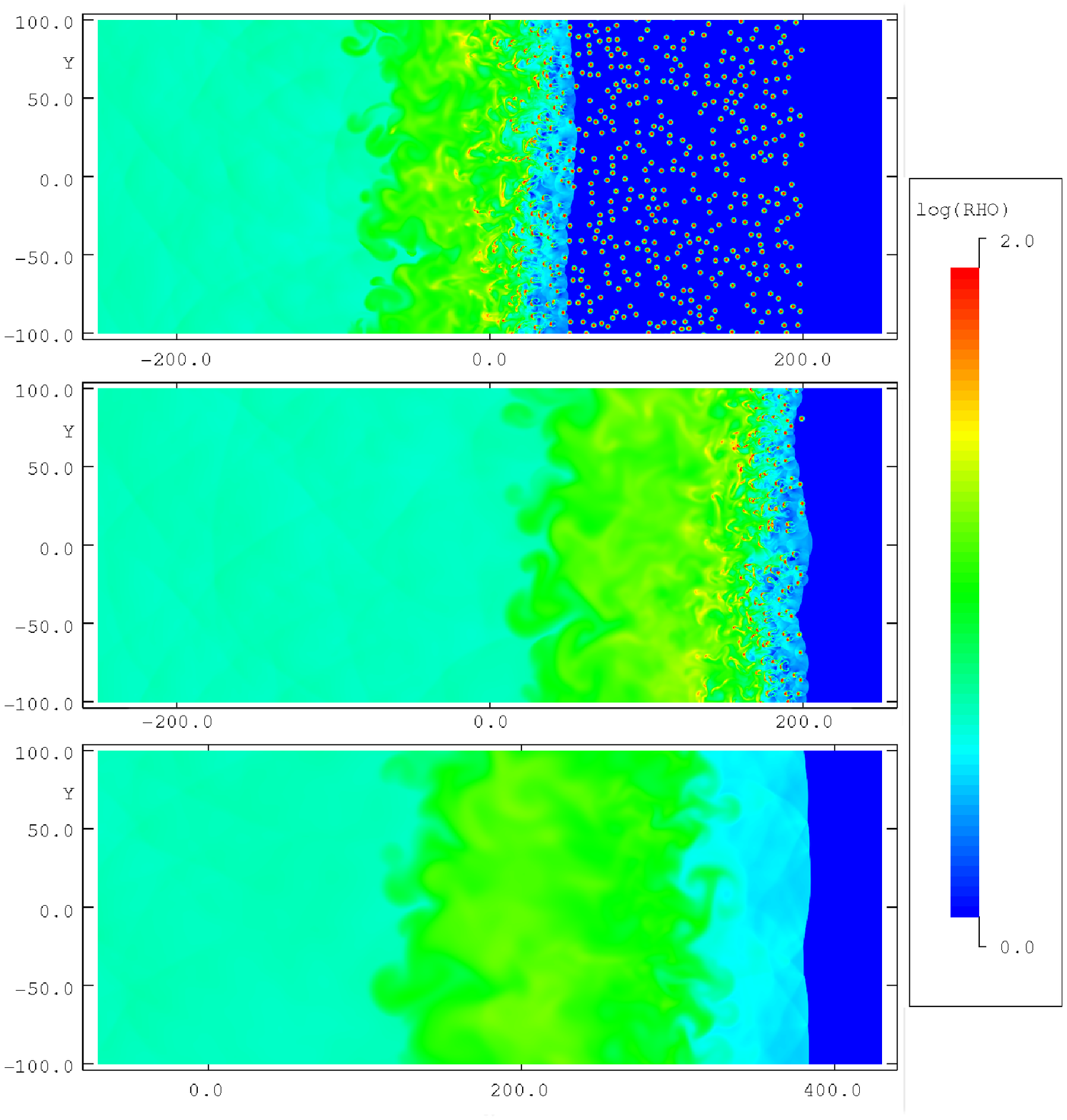}
\caption{As Fig.~\ref{fig:m3w6_early} but at $t=36.6\,t_{\rm cc}$
  (top), $t=61.0\,t{\rm cc}$ (middle) and 
$t=85.5\,t_{\rm cc}$ (bottom).
 }
\label{fig:m3w6_late}
\end{figure*}

We first focus on simulations for a Mach 3 shock.
Figs.~\ref{fig:m3w6_early} and~\ref{fig:m3w6_late} show the time
evolution of the density distribution in simulation \emph{m3chi2MR4},
which contains 1959 clouds. The shock sweeps through the clumpy
region, propagating fastest through the channels between
clouds. Behind the shock, the overrun clouds are in various stages of
destruction, with the clouds closest to the shock being those which
are in the earliest stage of interaction and which, therefore, are the
most intact. Further behind the shock the clouds gradually lose their
identities as they are mixed into the
flow. Fig.~\ref{fig:m3w6_early} shows that a global bowshock moves
upstream into the post-shock flow. Although the bowshock disappears
out of view in Fig.~\ref{fig:m3w6_late}, it remains on the grid which
extends to $x = -450 $. Low density (and pressure) regions in the
post-shock flow are visible behind clouds as the flow rushes pass. The
global shock front is momentarily deformed by each cloud that it
encounters as it sweeps through the clumpy region. These local
deformaties in the shock front gradually accumulate into a distortion
of the shock front on larger length scales. The shock is most deformed
as it reaches the end of the clumpy region (see the middle panel of
Fig.~\ref{fig:m3w6_late}). The other striking feature of the
interaction is the formation of a dense shell in the post-shock flow
due to the addition of mass from the destruction of the clouds. This
is clearly seen in the panels in Fig.~\ref{fig:m3w6_late} and is the
most important large-scale structure formed in the global flow as a
result of the interaction.

The shock slows as it sweeps through the clumpy region and transfers
momentum to cloud material. When
the clumpy region is highly porous, the deceleration of the shock
front is minimal and the clouds gradually reach the velocity of the
postshock flow. This occurs if the global cloud cross section is small
(for instance, when the mass ratio is low (e.g., $MR=0.25$) and/or
when the cloud density contrast $\chi$ is high (e.g., $\chi=10^3$,
$MR=1$ as in simulation \emph{m3chi3MR1}).

\begin{figure*}
\includegraphics[width=\textwidth]{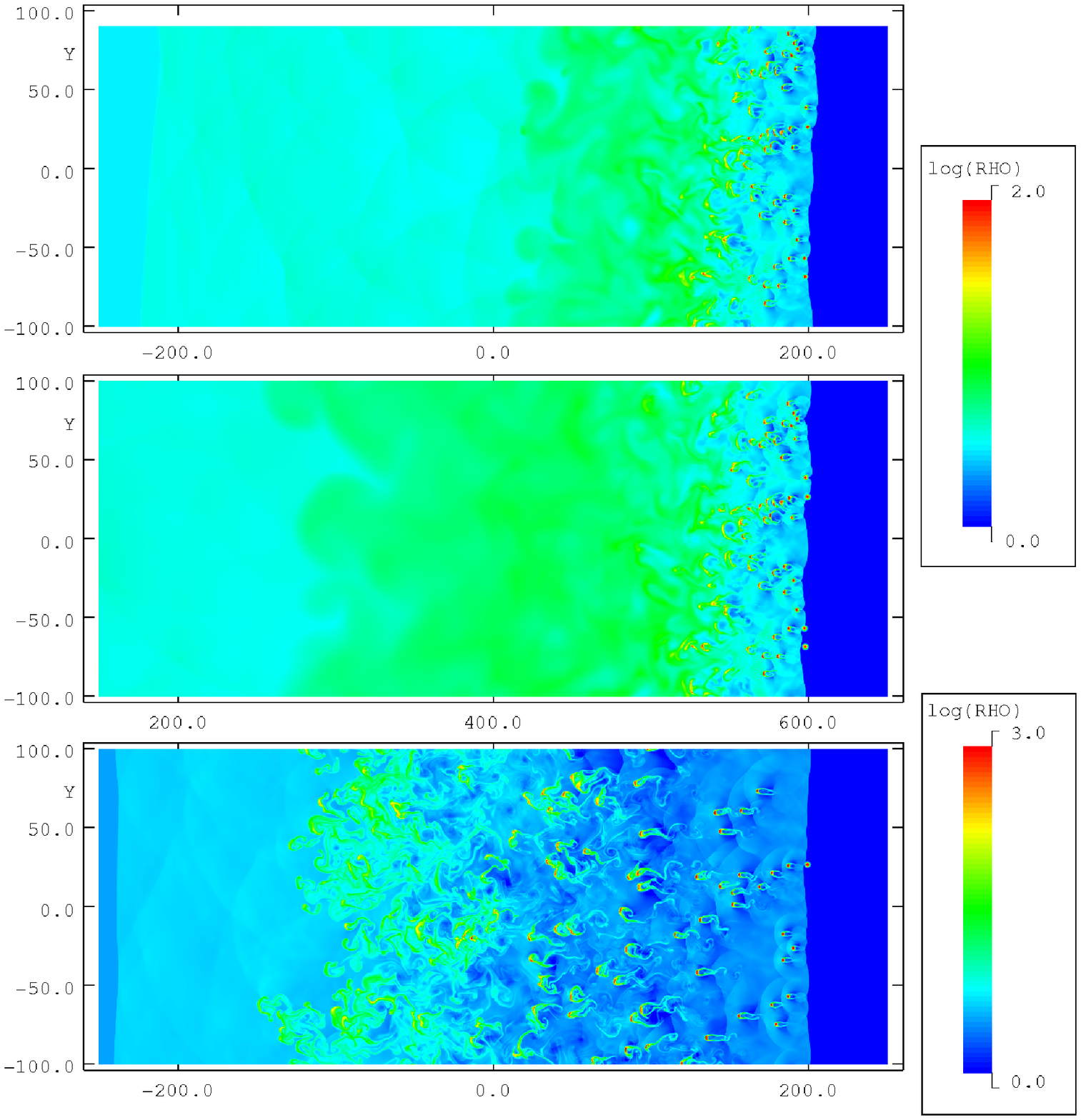}
\caption{Snapshots of logarithmic density for models \emph{m3chi2MR1} (top), \emph{m3chi2MR1\_double} (middle),
and \emph{m3chi3MR4} (bottom). The snapshots are taken as the
shock front is exiting the cloudy region which occurs at 
$t=46\,t_{\rm cc}$, $t=95\,t_{\rm cc}$ and $t=13.5\,t_{\rm cc}$ respectively. Note the different colour scale
of the bottom panel.  }
\label{fig:m3_differentW}
\end{figure*}

Fig.~\ref{fig:m3_differentW} shows snapshots of the density
distribution at the time that the shock exits the clumpy region in
models {\em m3chi2MR1} (top), {\em m3chi2MR1\_double} (middle), and
{\em m3chi3MR4} (bottom). All of these models have lower
number densities of clouds than model {\em m3chi2MR4}: in the
first two models it is because the value of $MR$ is reduced, while in
model {\em m3chi3MR4} it is because the density contrast of the clouds
has increased while the value of $MR$ is unchanged. As such, the 
clumpy region in each of these models is more porous than in model
{\em m3chi2MR4}, and the shock is able to sweep through without
such a significant reduction in its velocity. Comparison of
Figs.~\ref{fig:m3w6_early}, \ref{fig:m3w6_late}
and~\ref{fig:m3_differentW} also reveals significant
differences between these models in the structures of the flow through the clumpy region.

\subsubsection{Velocity behaviour}
\label{sec:velocity}
As the shock sweeps through a clumpy region, it causes clouds in block 1 to accelerate first; then clouds in blocks 2, 3
and 4 accelerate. This transfer of momentum to the clouds inevitably causes the
shock and the post-shock flow to slow.
Figs.~\ref{fig:vx_wide69} and~\ref{fig:vx_wd5} show the time
evolution of the average velocity of cloud material within particular blocks of clouds
(e.g., as delineated in the top panel of Fig.~\ref{fig:m3w6_early})
normalized to the intercloud sound speed.

Fig.~\ref{fig:vx_wide69} shows that the shock front slows down to
about 50\% of its initial velocity (though it remains supersonic). In
this simulation we find that the first 3 blocks of clouds accelerate
up to $1.25\,c_1$ (where $c_1$ is the sound speed in the intercloud ambient
medium). However, by this point the post-shock flow immediately after
the shock has decelerated to marginally below $1\,c_1$. Hence, material
stripped from the upstream clouds pushes against the downstream
clouds, compressing the clumpy region.

The shock front advances at roughly constant velocity after having
reached a minimum while in the clumpy region.  However, as the shock
front leaves the clumpy region it reaccelerates at a roughly constant
rate until it reaches a velocity somewhat smaller than the initial
shock velocity.  After that it accelerates very slowly, and
asymptotically reaches its initial velocity (corresponding to Mach
3).

The final block of clouds is not held back by further clouds
downstream and therefore does not stop accelerating at the same
velocity as the other blocks. Instead it accelerates until it reaches
the velocity of the post-shock flow after the shock stops
reaccelerating. Blocks 3, 2 and 1 repeat this behaviour in that order.

Figs.~\ref{fig:vx_wd5}a) and b) show the velocities obtained in
simulations {\em m3chi2MR1} and {\em m3chi2MR1\_double}, for which the mass
ratio of cloud to intercloud material in the clumpy region is unity
(the number density of clouds is therefore $4 \times$ lower than in
model {\em m3chi2MR4}). Model {\em m3chi2MR1\_double} has a
cloud distribution that is identical to that of 
model \emph{m3chi2MR1}, but the distribution is repeated once to
create a clumpy region that is twice as deep. In both cases the clumpy
region is more porous and therefore, the reductions in the shock and
post-shock velocities are not as severe as those seen in model {\em
  m3chi2MR4}. For this reason, the clouds are also accelerated
initially to a higher velocity ($\approx 1.7 c_{1}$) than in model
{\em m3chi2MR4}.

As previously stated, the clumpy region can also be made more porous
by increasing $\chi$ for a fixed $MR$. Fig.~\ref{fig:vx_wd5}c)
shows the velocity profiles obtained from
model \emph{m3chi3MR4}. As expected, the shock does
not decelerate as much as in model {\em m3chi2MR4}. However, model
{\em m3chi3MR4} behaves differently in another way: because each cloud is more resistant to the
flow, the clouds are not completely destroyed by the time the shock
leaves the clumpy region (at $t \approx 13\,t_{\rm cc}$) when
the collective cloud material from each block is still
accelerating. In turn, due to continuing mass-loading of the flow, the
shock continues to decelerate {\em after} it has left the clumpy
region. In fact, the shock front decelerates until $t \approx
36\,t_{\rm cc}$, remains at constant velocity until $t \approx
44\,t_{\rm cc}$ and reaccelerates thereafter.

\subsubsection{Stages of interaction}
\label{sec:stages}
The behaviour of the simulations noted in the previous section allows
the identification of 4 distinct phases in the evolution of the shock
front.  Firstly, there is a shock \emph{deceleration} phase as the
shock enters the clumpy region.  This may be followed by a \emph{steady}
phase, when the shock front moves at constant velocity.  The third
stage is a \emph{reacceleration} phase during which the shock front
accelerates (at a fairly steady rate) into the homogeneous ambient
medium. This stage always starts after the shock has traversed through
the whole of the cloudy region, but not necessarily immediately after.
The \emph{final} stage begins once the shock's acceleration slows. At
this point the shock propagates with a velocity slightly slower than its
initial velocity, but is continuing to accelerate very slowly, due to
the constant velocity inflow trying to return the shock to its initial
velocity.

\begin{figure}
  \begin{center}
    \begin{tabular}{c}
\psfig{figure=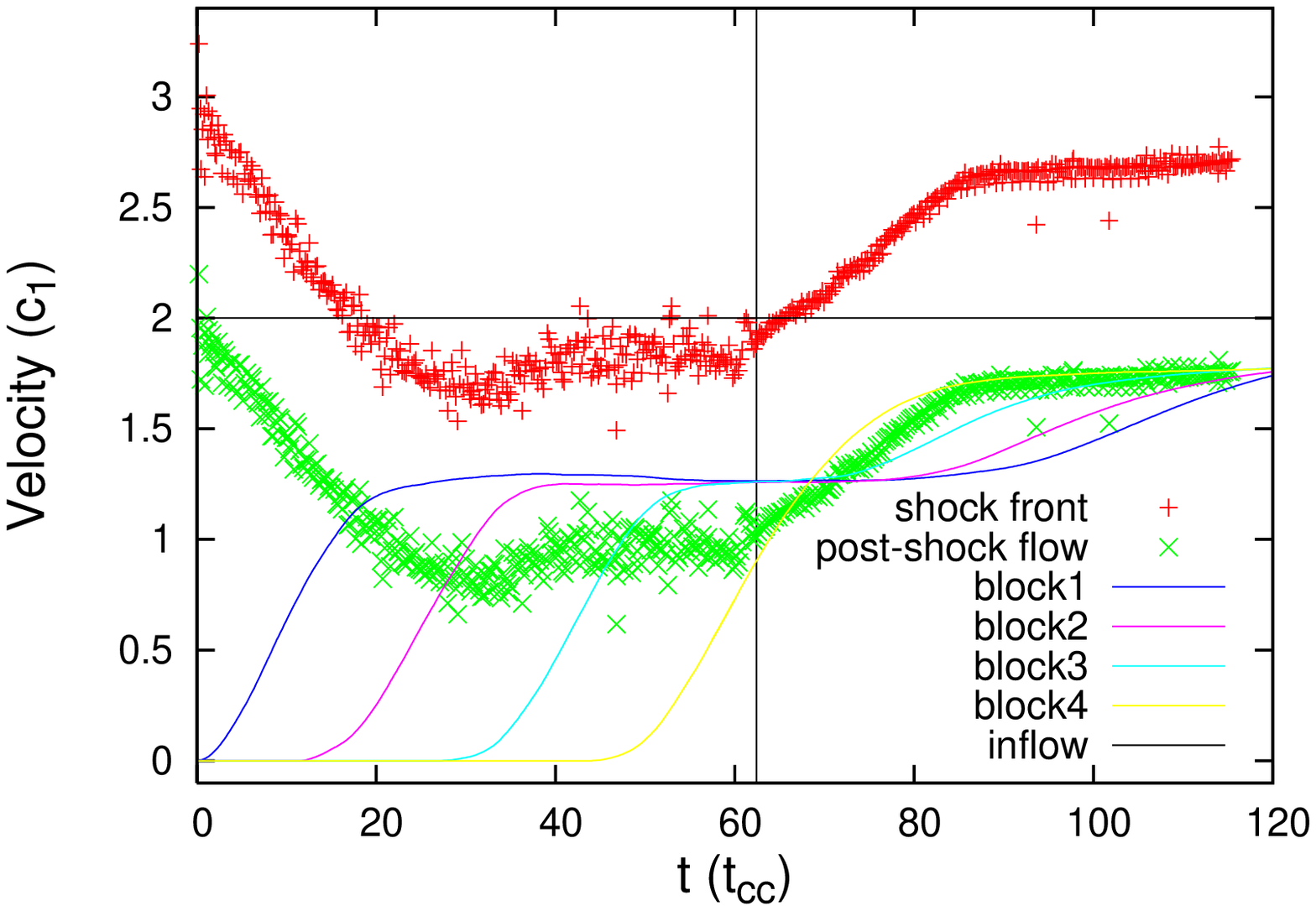, angle=-90, width=6.5cm}\\
     \end{tabular}
\caption[]{ The time evolution of the mean $x$-velocity for cloud
  material for each different block in simulation
\emph{m3chi2MR4}. Also shown are the average velocity of the shock 
front and the velocity immediately behind the shock 
front. The vertical line indicates the time when the shock leaves the
clumpy region, while the horizontal line indicates the initial speed of
the post-shock flow.}
\label{fig:vx_wide69}
\end{center}
\end{figure}

\begin{figure}
  \begin{center}
    \begin{tabular}{c}
\psfig{figure=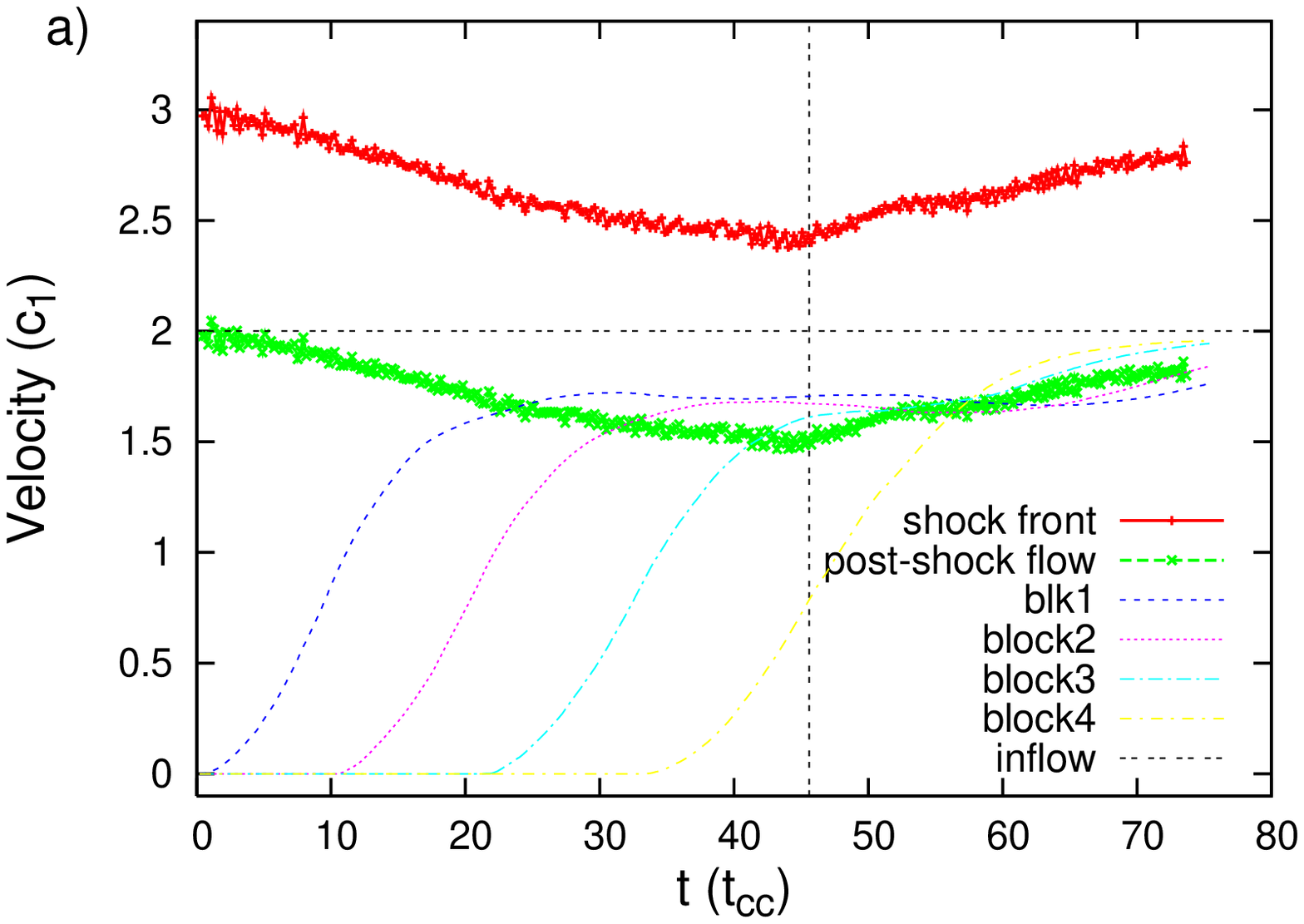, angle=-90, width=6.5cm}\\
\psfig{figure=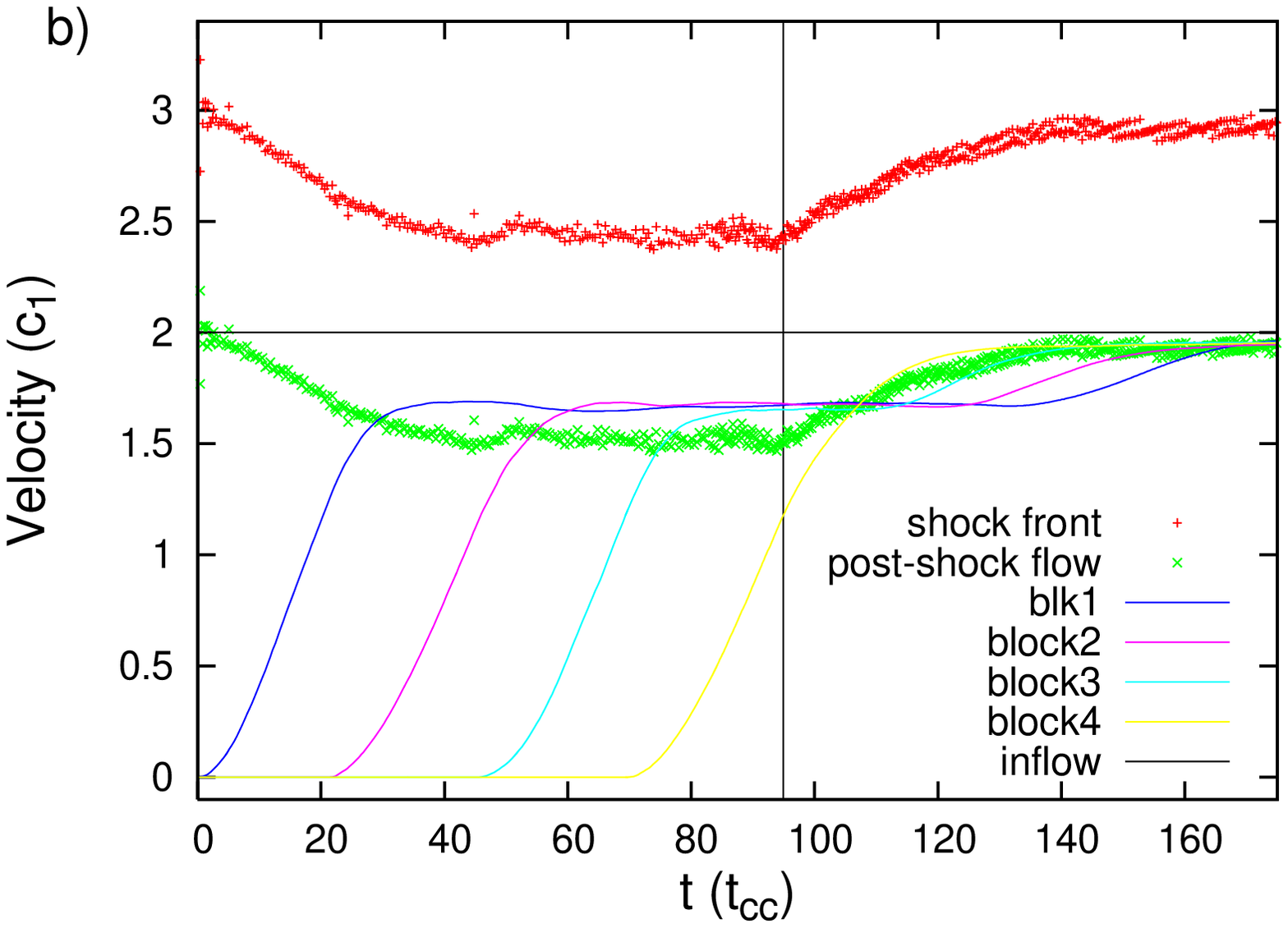, angle=-90, width=6.5cm} \\
\psfig{figure=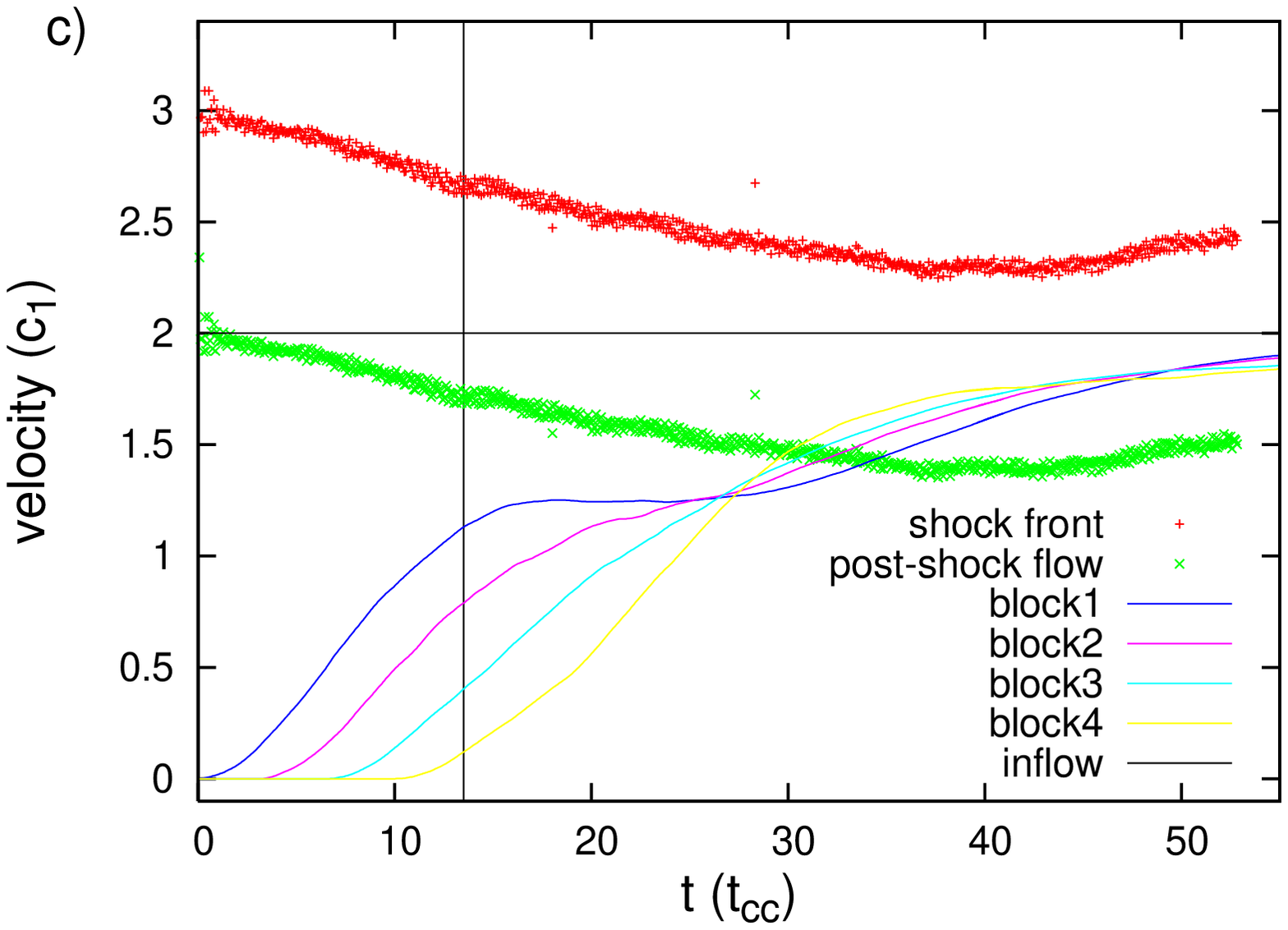, angle=-90, width=6.5cm} \\
      \end{tabular}
\caption[]{As Fig.~\ref{fig:vx_wide69} but for simulations
a) \emph{m3chi2MR1}, b) \emph{m3chi2MR1\_double} and c) \emph{m3chi3MR4}. }
\label{fig:vx_wd5}
\end{center}
\end{figure}

The number and timings of the stages is model specific. In models
for which the lifetime of the clouds is less than the crossing time of the
shock through the clumpy region, the steady phase ends when the shock
leaves the clumpy region (see, \eg, simulation \emph{m3chi2MR4} in
Fig.~\ref{fig:vx_wide69}). In models for which significant mass
loading of the flow continues after the shock leaves the clumpy region
the end of the steady phase is delayed (see, \eg, \emph{m3chi3MR4} in
Fig.~\ref{fig:vx_wd5}c).

The steady stage is best seen in models \emph{m3chi2MR4} and
\emph{m3chi2MR1\_double}, and to some extent it is also visible in
model \emph{m3chi3MR4}. These are the simulations with the most mass
in the clouds. Models \emph{chi2MR4} and \emph{chi3MR4} have the
highest cloud to intercloud mass ratios of 4.  
In fact, the onset of a steady stage seems to intimately depend on
the formation of a dense shell (see Sec.~\ref{sec:shell}).

In contrast, model \emph{m3chi2MR1} (Fig.~\ref{fig:vx_wd5}a) does not
achieve a steady state. Instead the shock velocity evolves from the
deceleration phase immediately into the reacceleration phase. One
might expect that there is a better chance for a steady stage to occur
if the clumpy region is deep. Fig.~\ref{fig:vx_wd5}b
indeed shows that a steady state occurs in model {\em
  m3chi2MR1\_double} where the clumpy region is twice as deep as in
model {\em m3chi2MR1}. The onset of the steady stage occurs when the shock is
approximately halfway through the clumpy region, and all 4 stages
can now be distinguished as in model \emph{m3chi2MR4}.

For all of the $\chi=10^2$ models, 
the reacceleration stage begins the moment the shock leaves the clumpy
region, but for the $\chi=10^3$ runs the onset of this stage shows more 
complicated behaviour. Because clouds with a higher density contrast evolve
more slowly, significant evolution of the global flow can occur after the
shock front leaves the clumpy region. Fig.~\ref{fig:vx_wd5}c shows
that this occurs in model {\em m3chi3MR4}, as the
continued deceleration of the shock is evident. At some later time, presumably
when the clouds are finally fully entrained into the flow, the shock
starts to accelerate. There is also a hint of a steady stage in model
\emph{m3chi3MR4}, but examination of the velocity graphs reveals that
it occurs at a much later evolutionary stage than in model
\emph{m3chi2MR4} (as seen, \eg, when compared to the velocity
evolution of material mixed in from the various blocks). Given that
there are about $10\times$ fewer clouds in model \emph{m3chi3MR4} than
in \emph{m3chi2MR4} this can be explained by the relative timescales
involved in the shell formation which is discussed next.

\subsubsection{Shell formation and evolution}
\label{sec:shell}
Figs.~\ref{fig:m3w6_early} and~\ref{fig:m3w6_late} show that 
in model \emph{m3chi2MR4}, a shell forms in the post-shock flow
at $t=16.2\,t_{\rm cc}$. It is fully developed by $36.6\,t_{\rm cc}$. At
$t=36.6\,t_{\rm cc}$ 7 distinct regions can be distinguished.  A
uniform ambient medium (without any clouds) lies at the right edge of
the figure, while at the far left of the grid (off the figure in this
case) lies the original post-shock flow specified by inflow boundary
conditions at the upstream edge.  A further 5 regions (from right to
left) exist inbetween these other two.  
A region of unshocked clouds lies in the range $50 \ltsimm
x < 200$. Shocked clouds embedded in a relatively low density
postshock flow lie in the region $30 \ltsimm x \ltsimm 50$.  Shocked
clouds are being overrun by a denser shell in the region $-20 \ltsimm
x \ltsimm 30$, though individual cores are still visible. This shell
becomes gradually more uniform towards $x \approx -70$.  The upstream
edge of this dense shell is fairly distinct from the less dense gas
that it abuts (at $x \approx -70$). The latter is gas in the
original supersonic post-shock flow which has shocked against the
clumpy region. This final region is bounded by a bow shock which lies
off the left edge of the figure.

\begin{figure}
\begin{center}
\psfig{figure=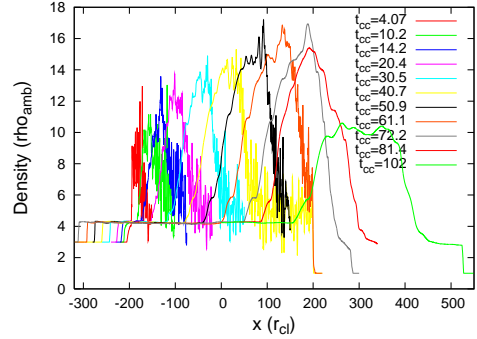, angle=-90, width=65mm}
\caption{The density, volume averaged over $y$, as a function of $x$
in simulation \emph{M3chi2MR4}, at different times. The shock exits
the clumpy region at $t \approx 60\,t_{\rm cc}$.}
\label{fig:1D_m3w6_rho}
\end{center}
\end{figure}

Fig.~\ref{fig:1D_m3w6_rho} 
shows the evolution of the density in the shell. Initially the average
postshock density in the clumpy region is only about twice that of the
shocked intercloud gas. However, as the clouds are destroyed and their
mass mixes into the global flow, the density in this region increases.
In model {\em m3chi2MR4}, a dense region, downstream of the region where
distinct clouds are still visible, has formed by about the time that
the shock reaches its minimum velocity. This is the ``shell''. The
compression of the shell progresses a little further, until a state
in which the maximum density is steady in time is reached. This is
reinforced by the simulation (model \emph{m3chi2MR1\_double}) for a
wide clumpy region,
which shows that the shell only becomes wider with time after this
point. Once there are no more clouds blocking the path of the shell, it
starts to expand from its leading edge, and the maximum density within
it drops. This is seen clearly from Fig.~\ref{fig:1D_m3w6_rho},
and is preceded by the shock reaching the edge of the clumpy region.
Simulations with $\chi=10^{3}$ behave slightly differently - in these
the density of the shell continues to increase after the shock has
left the clumpy region due to the continued ablation of material from
the longer-lived clouds.

\subsubsection{Mach number profile}
\begin{figure}
\begin{center}
\psfig{figure=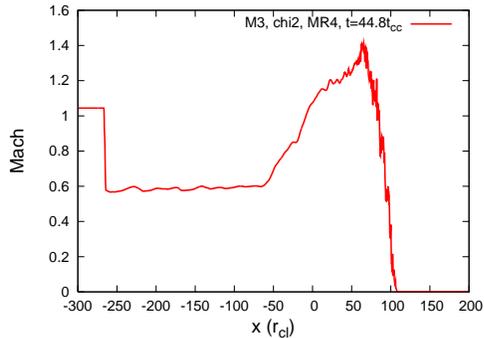, angle=-90,  width=65mm}
\caption{The local Mach number, volume averaged for all values of $y$, as a function of $x$
in simulation \emph{M3chi2MR4}, at $t=44.8\,t_{\rm
  cc}$. The shock is at $x \approx 108$ and the bow shock is at $x
\approx -265$.}
\label{fig:machprofile}
\end{center}
\end{figure}

Fig.~\ref{fig:machprofile} shows the $y$-averaged Mach number profile
of the flow in simulation \emph{M3chi2MR4} at $t=44.8\,t_{\rm
  cc}$. The undisturbed post-shock flow is mildly
supersonic ($M \approx 1.04$), as seen at the far left of the
panel. The shock at this time is just downstream of $x = 100$. 

In the region between $70 \ltsimm x \ltsimm 108$ clouds are overrun by
the flow and accelerated by the gas streaming past, so that the Mach
number of the flow increases as $x$ decreases. While the clouds remain
identifiable as distinct entities, their interiors remain colder than
the shocked intercloud flow, as initially the dense cloud is in pressure equilibrium with
its less dense surroundings, so that its temperature is lower than that of the intercloud medium
by a factor of $\chi$. Together these effects ensure that
the Mach number of the flow eventually exceeds the Mach number in the
undisturbed post-shock flow. The local Mach number reaches a peak of
about 1.4, at a location corresponding to the downstream edge of the
shell (which occupies a region between $-50 \ltsimm x \ltsimm 70$).

As the clouds begin to lose their individual identities their material
is heated as it is mixed into the surrounding flow. This increases the
sound speed of this material, and reduces the overall Mach number of
the flow. The Mach number continues to decline with decreasing $x$
until the cloud material is fully mixed (which occurs at the upstream
edge of the shell in this simulation).  The region of constant Mach
number ($M\approx0.6$) between the upstream edge of the shell (at
$x\approx -50$) and the bow shock at the head of the clumpy region (at
$x\approx -265$) is the gas in the original post-shock flow which has
shocked against the clumpy region.

\subsection{Dependence on shock Mach number}
\begin{figure*}
\includegraphics[width=\textwidth]{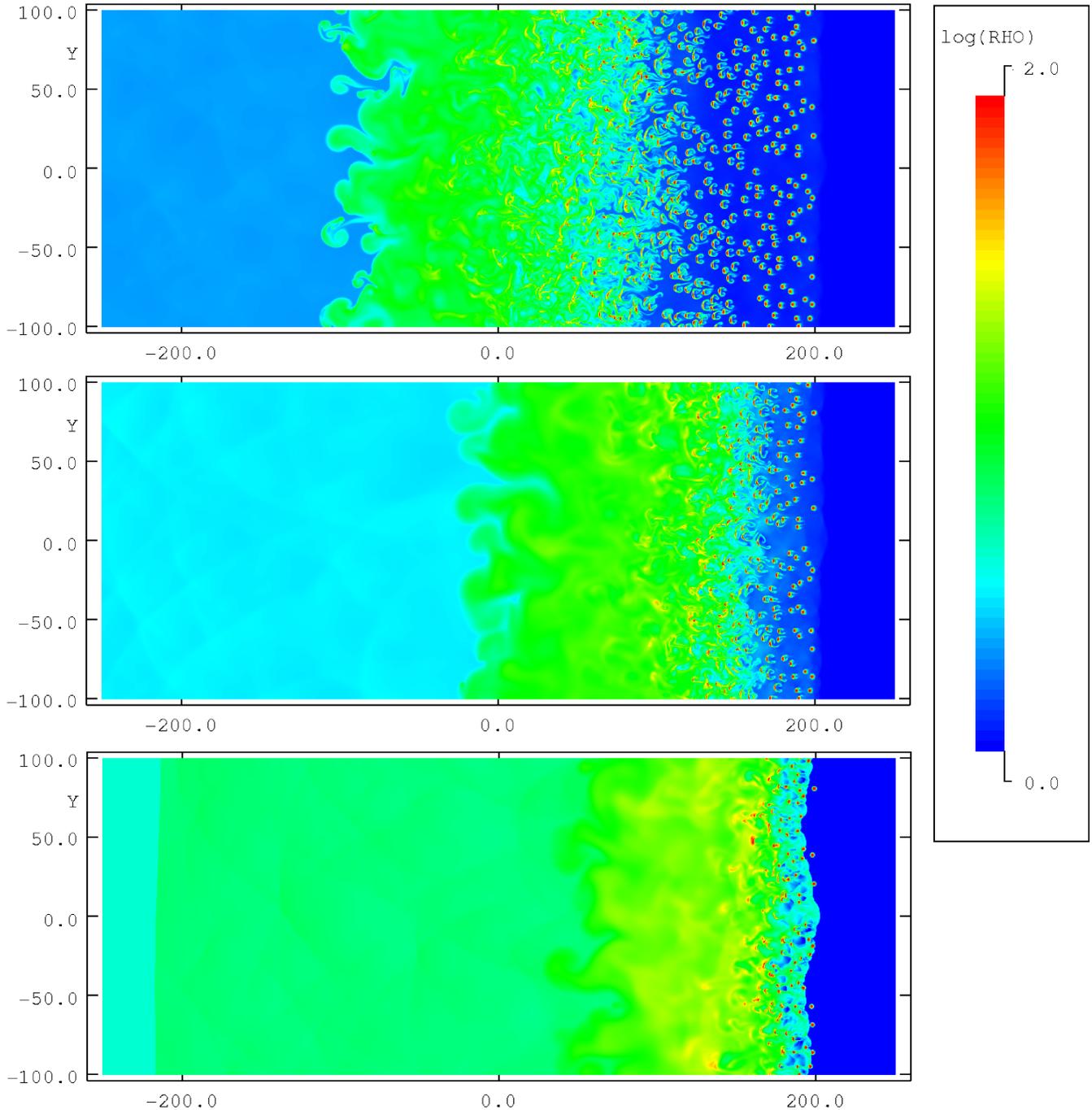}
\caption{Snapshot of logarithmic density for models \emph{m1.5chi2MR4} (top), \emph{m2chi2MR4} (middle)
and \emph{m10chi2MR4} (bottom). The snapshots are taken as the
shock front is exiting the cloudy region which occurs at 
$t=53\,t_{\rm cc}$, $t=62\,t_{\rm cc}$ and $t=58\,t_{\rm cc}$
respectively. Relative to its initial speed, the $M=2$ shock is slowed
the most.
 }
\label{fig:w6_differentM}
\end{figure*}

Fig.~\ref{fig:w6_differentM} shows density snapshots when the shock
front is exiting the clumpy region for models {\em m1.5chi2MR4}, {\em
  m2chi2MR4} and {\em m10chi2MR4}. It is clear that a slower shock
produces less compression, and that the shell which then forms is
wider. In addition, weak shocks take longer to destroy clouds, 
so the shell forms further from the shock
front. Perhaps most significantly, as the shock decelerates
during its passage through the clumpy region it can slow so much that
it decays into a wave which advances at the intercloud sound
speed. This is seen in both the $M=1.5$ and $M=2$ simulations for which
Fig.~\ref{fig:w6_differentM} reveals very weak density jumps at the
edges of the clumpy regions.

\begin{figure}
  \begin{center}
\psfig{figure=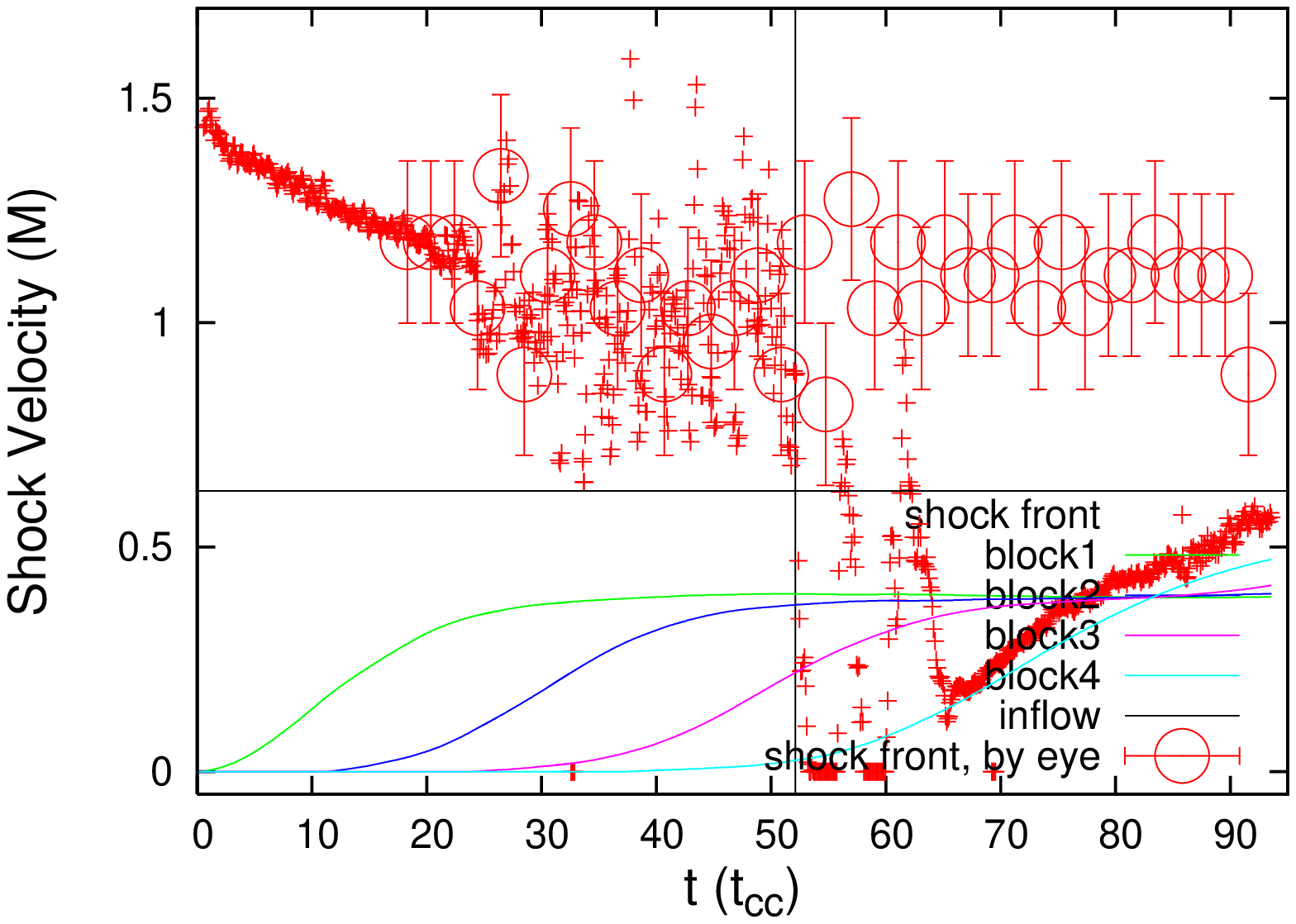, angle=-90, width=6.5cm}
\psfig{figure=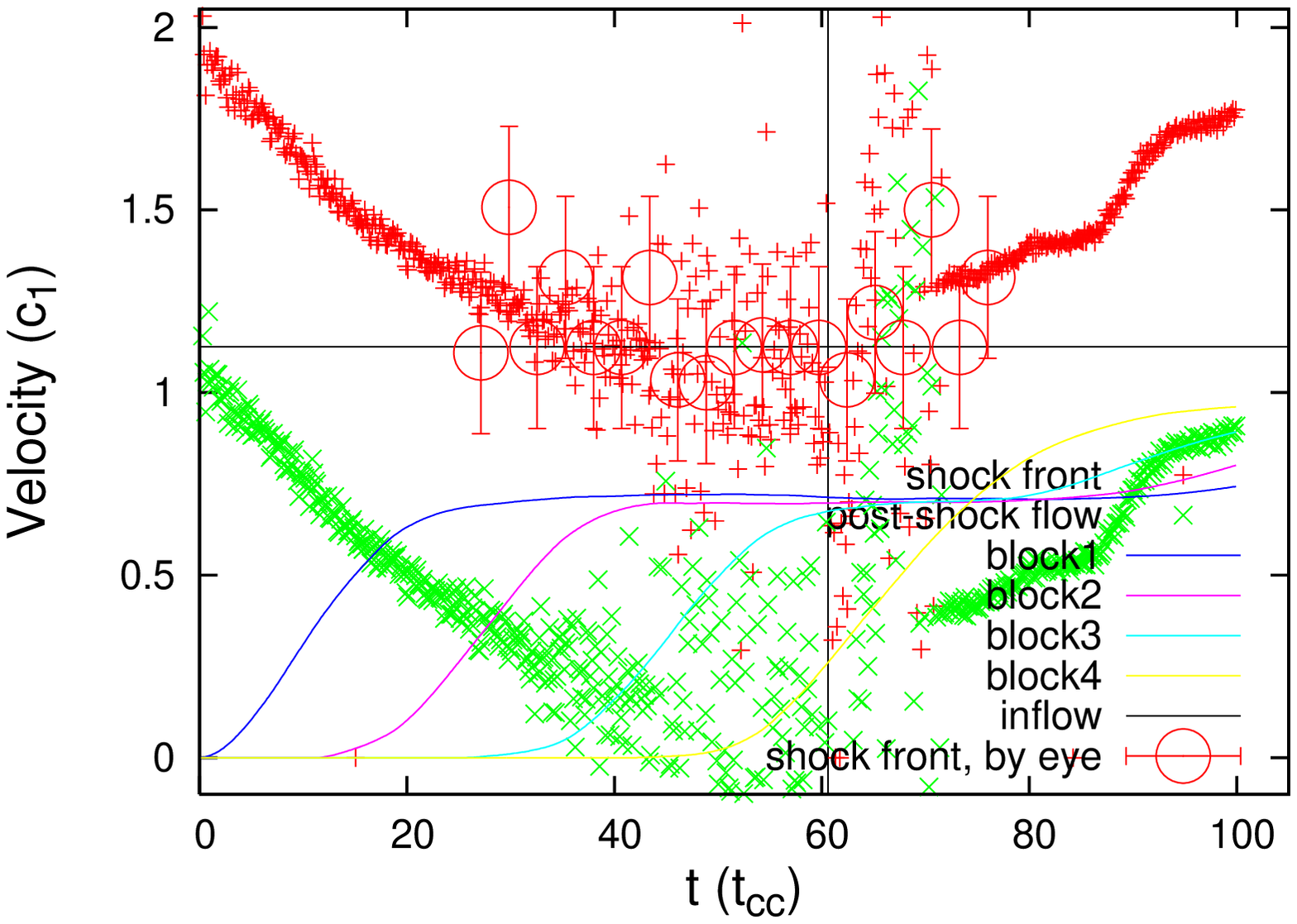, angle=-90, width=6.5cm}
\caption[]{ The time evolution of the $x$ component of the velocity averaged over all values of $y$,
for different regions in simulations 
\emph{m1.5chi2MR4} (top) and {\em m2chi2MR4} (bottom). Also shown are the average shock 
front velocities. The shock velocity varies when parts of the shock front become 
subsonic. Vertical lines indicate the time when the shock leaves the clumpy region. 
}
\label{fig:vx_wide6_slow}
\end{center}
\end{figure}

Fig.~\ref{fig:vx_wide6_slow} displays the $x$ component of the velocity
averaged over all $y$ for the {\em m1.5chi2MR4} and {\em m2chi2MR4} simulations
normalized to the ambient intercloud sound speed. The points show the
velocity of the shock calculated by an algorithm which
searches for a pressure jump.  However, when the shock decays into a
wave the pressure jump largely disappears and this calculation
fails. The points with error bars then show the velocity of the
disturbance determined by measuring, by eye, the position of the 
leading edge of the disturbance. In the $M=2$ simulation the average velocity of the
shock/wave disturbance is close to transonic. Close examination
reveals that the shock appears fragmented and local regions of sonic
waves can be seen. The shock completely decays into a sonic wave in
the $M=1.5$ simulation.  These results are in harmony with
theoretical predictions based on a uniform density region 
suggesting that a supersonic to subsonic transition occurs at this mass ratio
if the initial shock Mach number is $M \sim 2$ (see
Sec.~\ref{sec:reduction} for further details).

\subsection{The reduction in the shock speed}
\label{sec:reduction}
\begin{figure}
\begin{center}
\psfig{figure=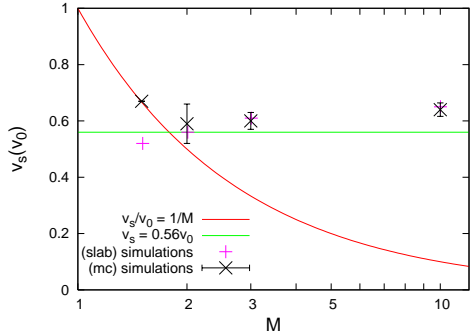, angle=-90, width=65mm}
\caption{The reduction in shock speed (black crosses, as measured during the
  ``steady'' phase as the shock propagates through the clumpy region)
  normalized to the initial shock speed. The reduction is plotted as a
  function of the initial shock Mach number, $M$, for simulations with
  $\chi=10^{2}$ and $MR = 4$. Also shown is the equivalent reduction
  in speed of a shock propagating through a uniform density
  enhancement (pink crosses - see Sec.~\ref{sec:uniform_rho}).
}
\label{fig:wide6_min_vel}
\end{center}
\end{figure}

Fig.~\ref{fig:wide6_min_vel} shows the minimum shock speed (normalized
to the initial shock speed) as a function of shock Mach number obtained in simulations with $MR =
4$ and $\chi=10^{2}$ (plotted as black crosses). The minimum shock speed in each case occurs
during a ``steady'' stage when the shock is moving through the
clumpy region. The error bars correspond to the $1\sigma$ spread in
measured speeds. We have already noted that in the $M=1.5$ simulation 
the shock slows to the point that the disturbance becomes a wave. In
this case the minimum speed is limited by the sound speed in the
intercloud medium, as seen in Fig.~\ref{fig:wide6_min_vel}. The
situation in the $M=2$ simulation is not so straightforward - a global
shock front is visible but there are local regions where a wave is
seen instead.

We can compare the measured shock speed to the expected speed of a shock transmitted
into a region of enhanced density. For this we make use of eq.~5.4 in
\citet{1994ApJ...420..213K}, for a shock interacting with a single cloud:

\begin{equation}
\label{eq:kmc}
v_{\rm s} = (F_{\rm st} F_{\rm c1})^{1/2} \chi^{-1/2} v_0.
\end{equation}

\noindent Here $F_{\rm st} \equiv P_3/P_1$, where $P_1$ is the pressure far
upstream and $P_3$ is the stagnation pressure (just upstream of the cloud),
and $F_{\rm c1} \equiv P_4/P_3$ is the ratio of the pressure just
behind the transmitted shock and the stagnation pressure. In the limit
of high Mach number and high density contrast,
\begin{equation}
\label{eq:fudge}
F_{\rm st} \simeq 1 + \frac{2.16}{1+6.55\chi^{-1/2}}. 
\end{equation}
To apply this expression to our multi-clump simulations we need 
an ``effective'' value for $\chi$ which accounts for the
inhomogeneity of the clumpy region. We choose this value to
be $\chi_{\rm eff} = MR +1$, which is the average density contrast of the region (total mass of the region divided
by the mass a same size region of ambient density would have). Furthermore, in simulations where the
clumpy region is replaced with a region of uniform density (see Sec.~\ref{sec:uniform_rho}), it is
clear that $P_4 = P_3 = P_2$, where $P_2$ is the pressure just after
the bow shock. We shall assume that this is also the case for our
multiple cloud simulations, and therefore adopt $F_{\rm c1} = 1$. For our clumpy
simulations with $MR=4$ and $\chi=10^{2}$ we therefore obtain
$\chi_{\rm eff} = 5$ and $F_{\rm st} = 1.55$, so that Eq.~\ref{eq:kmc}
gives $v_{\rm s} = 0.56 v_{0}$.  As can be seen, this is comparable to
the values shown in Fig.~\ref{fig:wide6_min_vel}, in which the $v_s = 0.56v_0$ line 
is plotted.

\subsection{Comparisons to a shock encountering a region of uniform
  density}
\label{sec:uniform_rho}
In the limit of an infinitely deep/thick cloud distribution we
expect the global behaviour of the shock to approach that of a shock
encountering a uniform medium of the same average density. This could
be similar to the steady state reached in some of our simulations.
Therefore, we have performed a number of additional calculations of a
shock encountering a uniform region of enhanced
density equal to the average density of a clumpy region with $MR=4$.
The minimum shock speed through this region is shown in 
Fig.~\ref{fig:wide6_min_vel} with pink crosses. As can be seen, the
agreement with the multi-cloud simulations is very good, with
significant deviation only at $M=1.5$. Clearly the minimum shock speed
obtained in our multi-cloud simulations is close to the
shock speed occuring in a comparable region of uniform density.

\begin{figure}
  \begin{center}
    \begin{tabular}{c}
      \psfig{figure=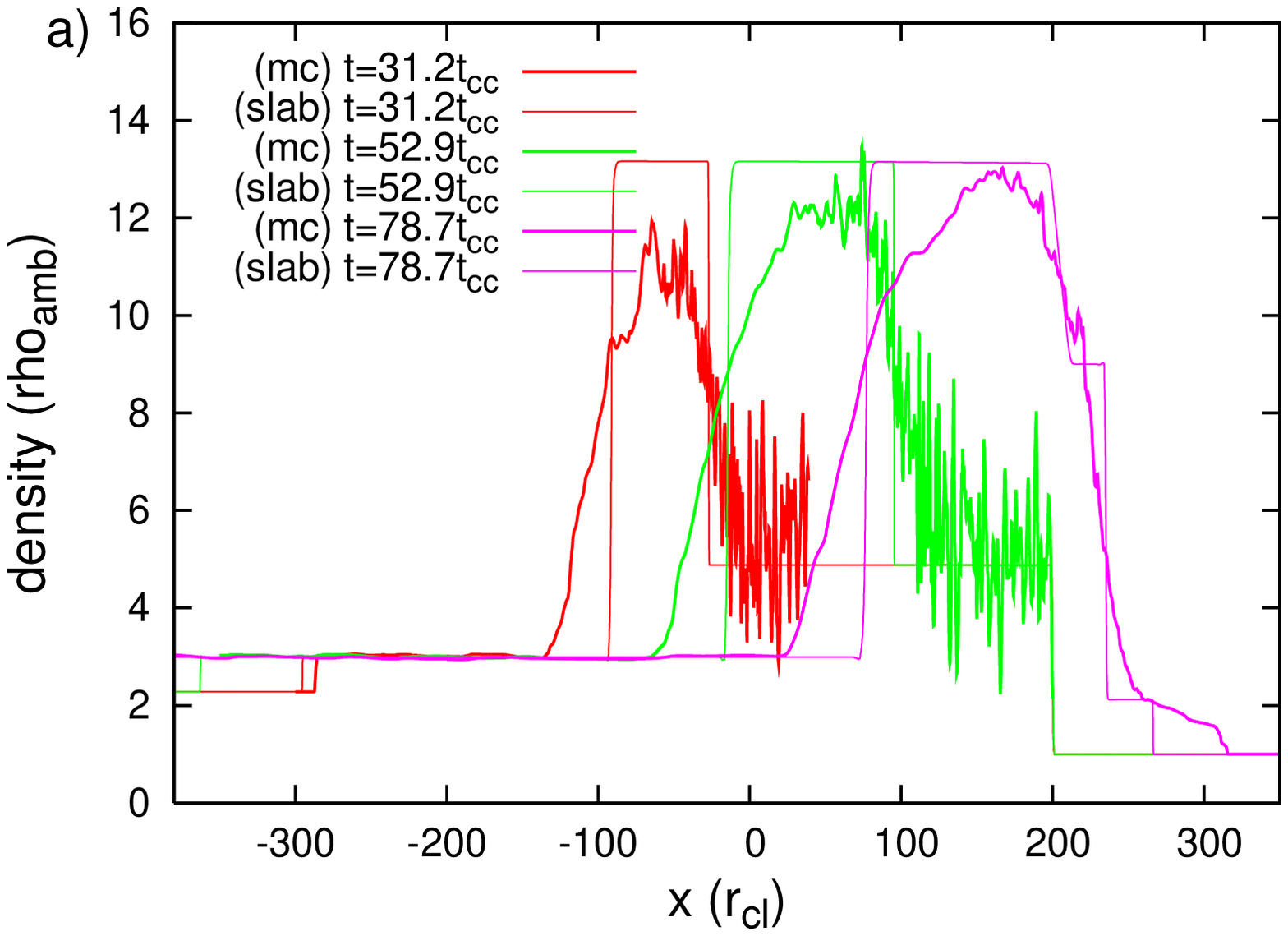, angle=-90, width=6.5cm}\\
      \psfig{figure=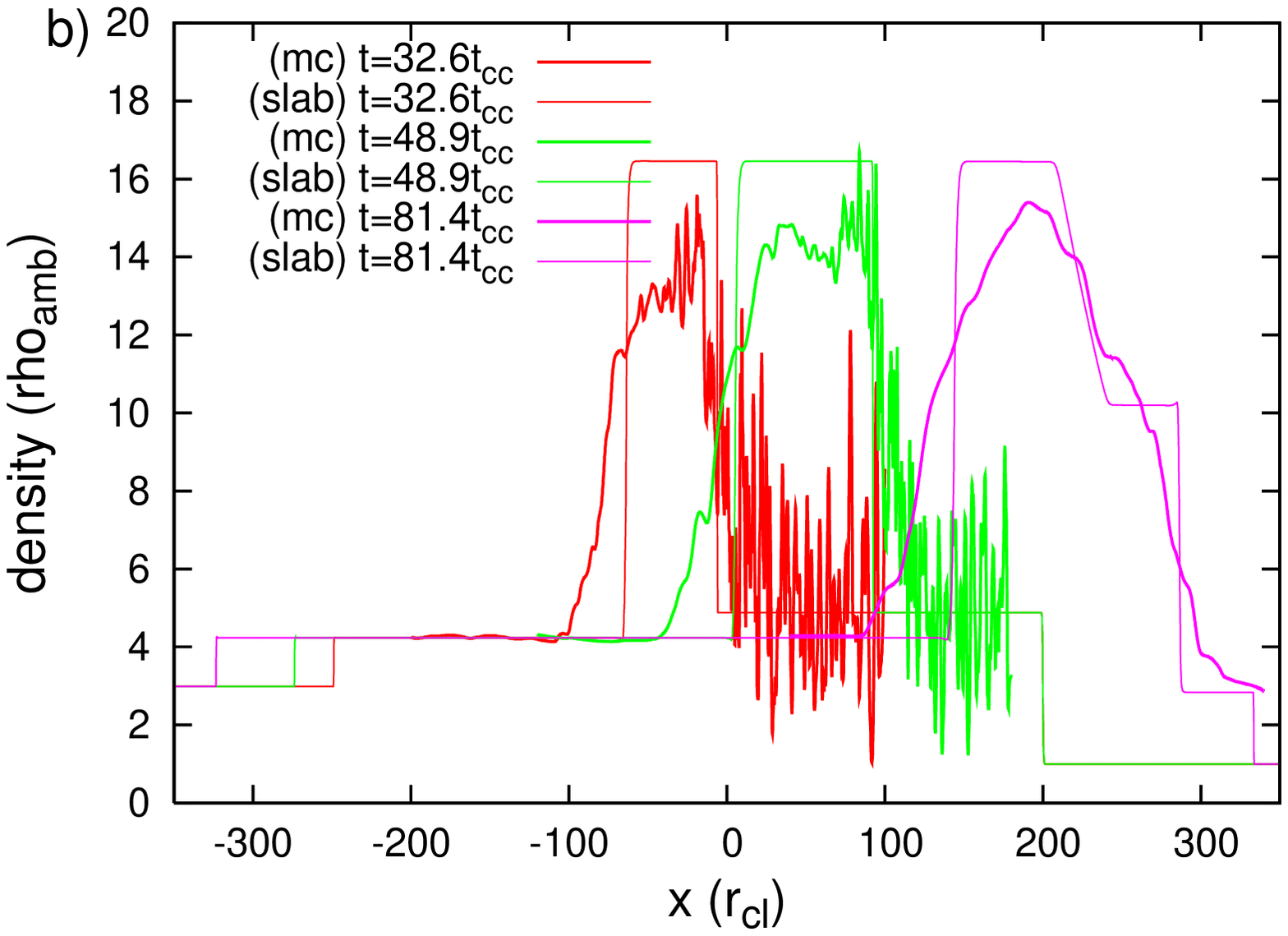, angle=-90, width=6.5cm} \\
      \psfig{figure=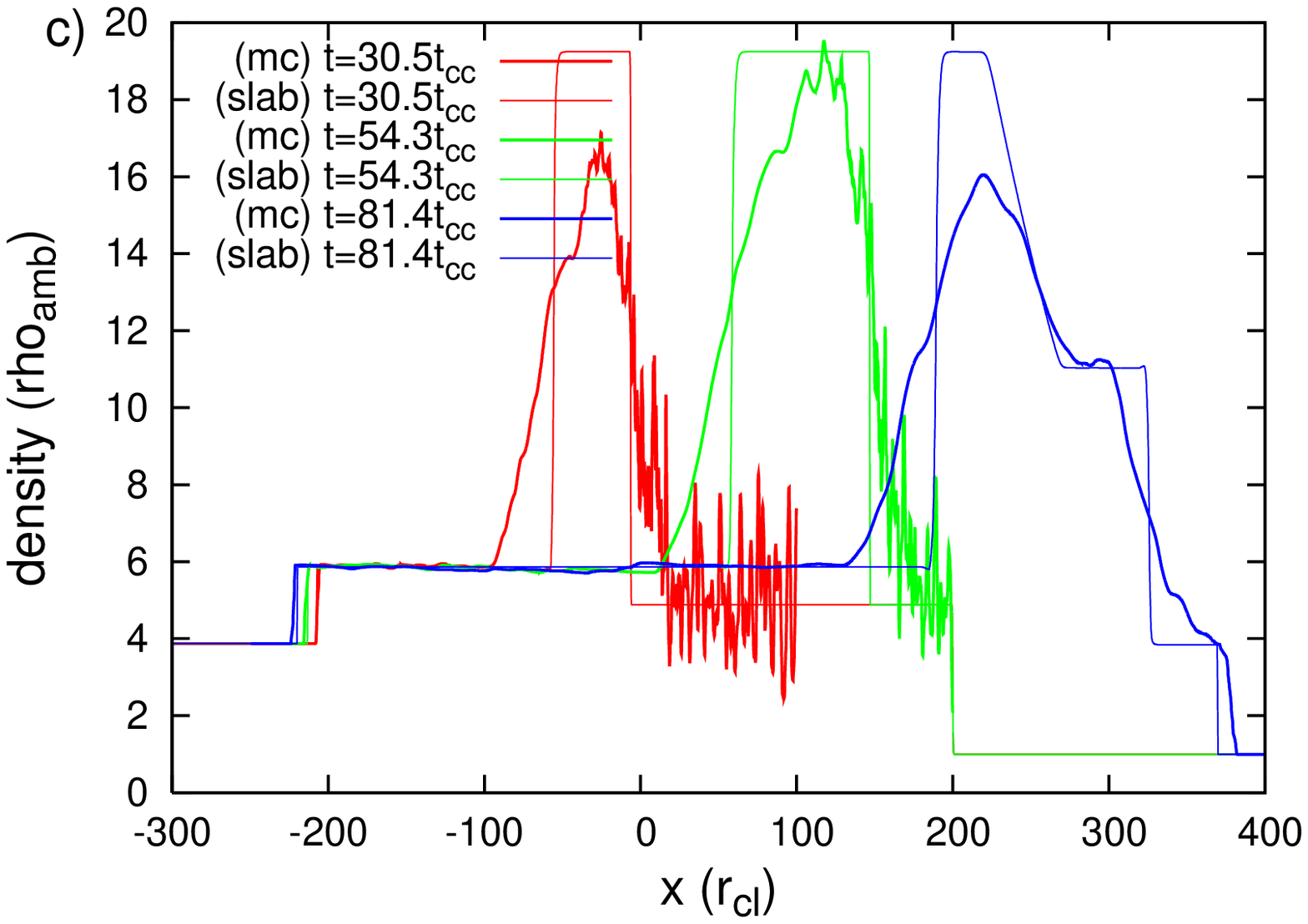, angle=-90, width=6.5cm} \\
      \end{tabular}
\caption[]{Snapshots of the $x$-dependance of the  density averaged over all values of $y$ 
(thick lines) from simulations a) \emph{m2chi2MR4}, b) \emph{m3chi2MR4} and
c) \emph{m10chi2MR4}. Also shown are density profiles
from simulations with uniform but corresponding density. In all cases the
shock exits the region of enhanced density (smooth or clumpy) at
$t\approx 60\,t_{\rm cc}$. 
}
\label{fig:shell_mc_slab}
\end{center}
\end{figure}

Fig.~\ref{fig:shell_mc_slab} compares the $x$-dependance of the density, averaged over all values of $y$,
given by simulations of a shock interacting with a uniform region
of enhanced density to analogous profiles obtained from our multiple cloud
simulations. One notices the similarity between
the profiles: the maximum density of the
shell and the reduction in density after the shock
exits the higher density/clumpy region are similar. Obviously, the
uniform region simulations yield distinct edges and slopes
corresponding to the shocks and rarefaction waves, whereas the
multiple cloud simulations give smoother profiles. The smoothing
length is of the order of $L_{\rm CD}$, which is the distance that a
cloud is displaced by the flow prior to its destruction
\citep{2002ApJ...576..832P}. For the same reason the width of the
densest part of the shell is narrower in the multiple cloud simulations.

The picture of a shock transmitted into a uniform density region is
different. The velocity jumps according to
Rankine-Hugoniot conditions. As the shock
front is moving faster than the postshock flow, the width, $w$ of
the compressed region is increasing at a rate given by:
\begin{equation}
\frac{dw}{dt} = v_s - v_{ps} = \left(\frac{3}{4M_u} + \frac{M_u}{4}\right)c_u.
\label{eq:dw}
\end{equation}

\noindent where $M_u$ is the Mach number of a shock transmitted into a
uniform medium and $c_u$ is the sound speed in that medium.
Eqs. \ref{eq:kmc} and \ref{eq:fudge} can be used to determine $M_u$,
but the approximation in Eq. \ref{eq:fudge} is only valid at high Mach
numbers and high density contrasts.  However, we can obtain an estimate
based on Eq.~\ref{eq:kmc}. As $F_{st} \geq 1$ and, because the shock
is steady, $F_c = 1$ \citep{1994ApJ...420..213K}, $M_u \geq M_0$ and
so $\frac{dw}{dt} \geq \frac{M_0c_1}{4}$. Furthermore, two limiting
cases can be found. In the limit of a weak shock $\frac{dw}{dt} \approx
c_u$; this is the minimum value, and it would be higher in all other
cases. In the limit of a strong shock and high $\chi$,
$\frac{dw}{dt} \approx 0.44M_0c_u$.

Fig. \ref{fig:shell_mc_slab} shows that the widths of shells in clumpy
regions correspond well to those in regions of equivalent uniform
density.  As such Eq. \ref{eq:dw} can be used, with $M_u$ either
determined from a corresponding uniform density case, or by replacing
it with $v_s / c_u$, with $c_u$ determined for the medium of
corresponding uniform density. The limiting cases would be different
though.  In the weak shock limit, the shock may decay into a sound
wave. In such a case $\frac{dw}{dt} \approx c_1$, but it could be
significantly less depending on how this affects individual cloud
lifetimes. In the limit of very high $\chi$ the upstream edge of the
shell takes a long time to reach $v_{ps}$ and is effectively
stationary, so $\frac{dw}{du} = v_s$.

\subsection{Individual cloud evolution}
\label{sec:individual}
We expect individual clouds to evolve differently when overrun by a
mass-loaded evolved shock which has been altered by the presence of
clouds further upstream than when they are overrun by a
steady planar shock. Section~\ref{sec:velocity} shows that the shock
Mach number and in turn the velocity of the post shock flow decreases
as the shock sweeps up cloud material. The lower velocity of the shock
should prolong the lifetime of a cloud, but this may be offset by the
increased density of the flow.  Furthermore, the postshock flow is
much more turbulent. This aids the development of cloud-destroying
instabilities \citep[see][]{2009MNRAS.394.1351P}. Finally, a region of higher
density follows some distance behind the shock. In the extreme case it
is a dense shell, but even if a shell does not fully form, remnants of
upstream clouds may well interact with clouds further
downstream. Therefore, it is difficult to predict whether
downstream clouds are destroyed more readily than their upstream
counterparts. However, we can use our simulation results to investigate this.

\begin{figure*}
\begin{center}
\psfig{figure=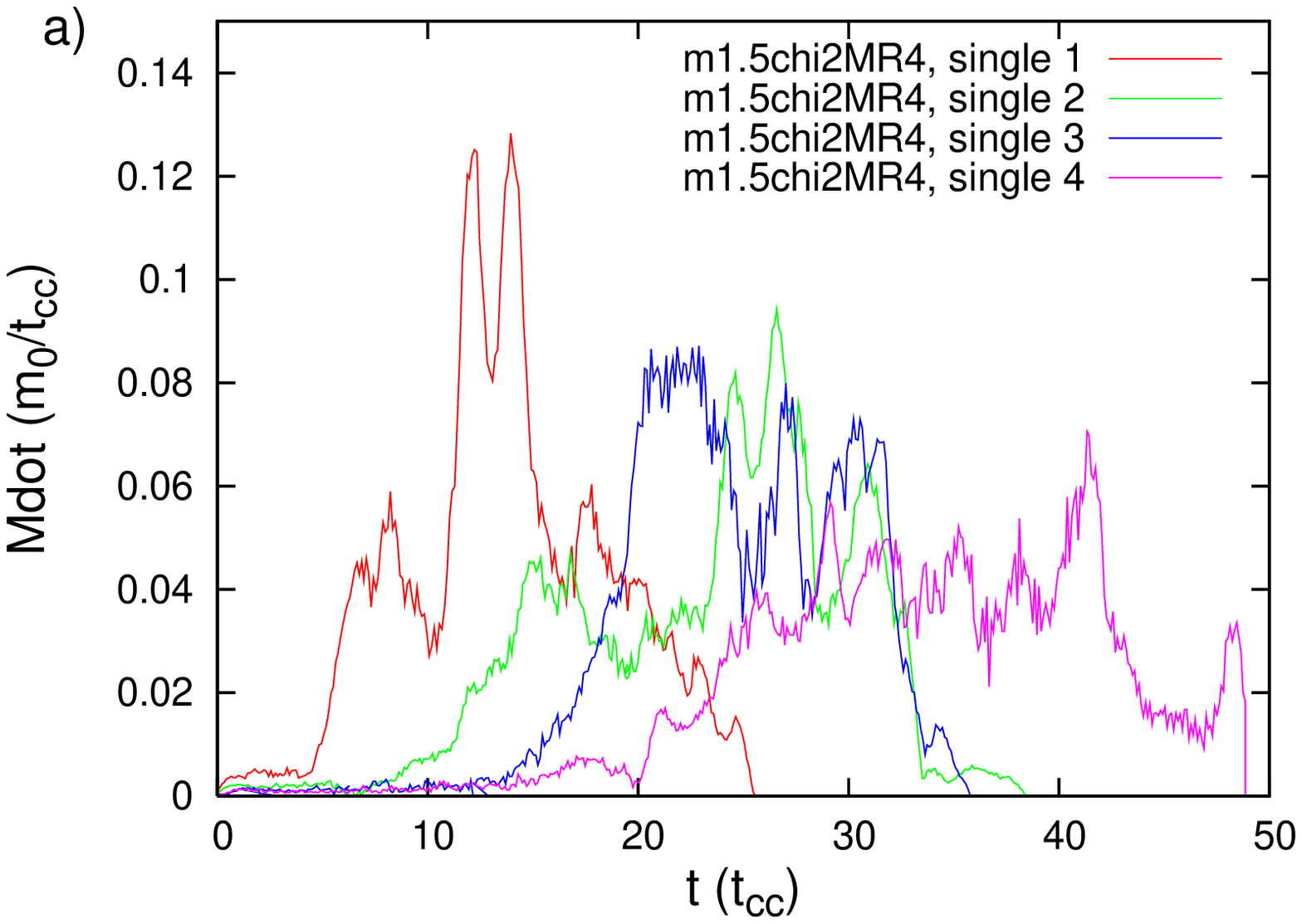, angle=-90, width=6.5cm}
\psfig{figure=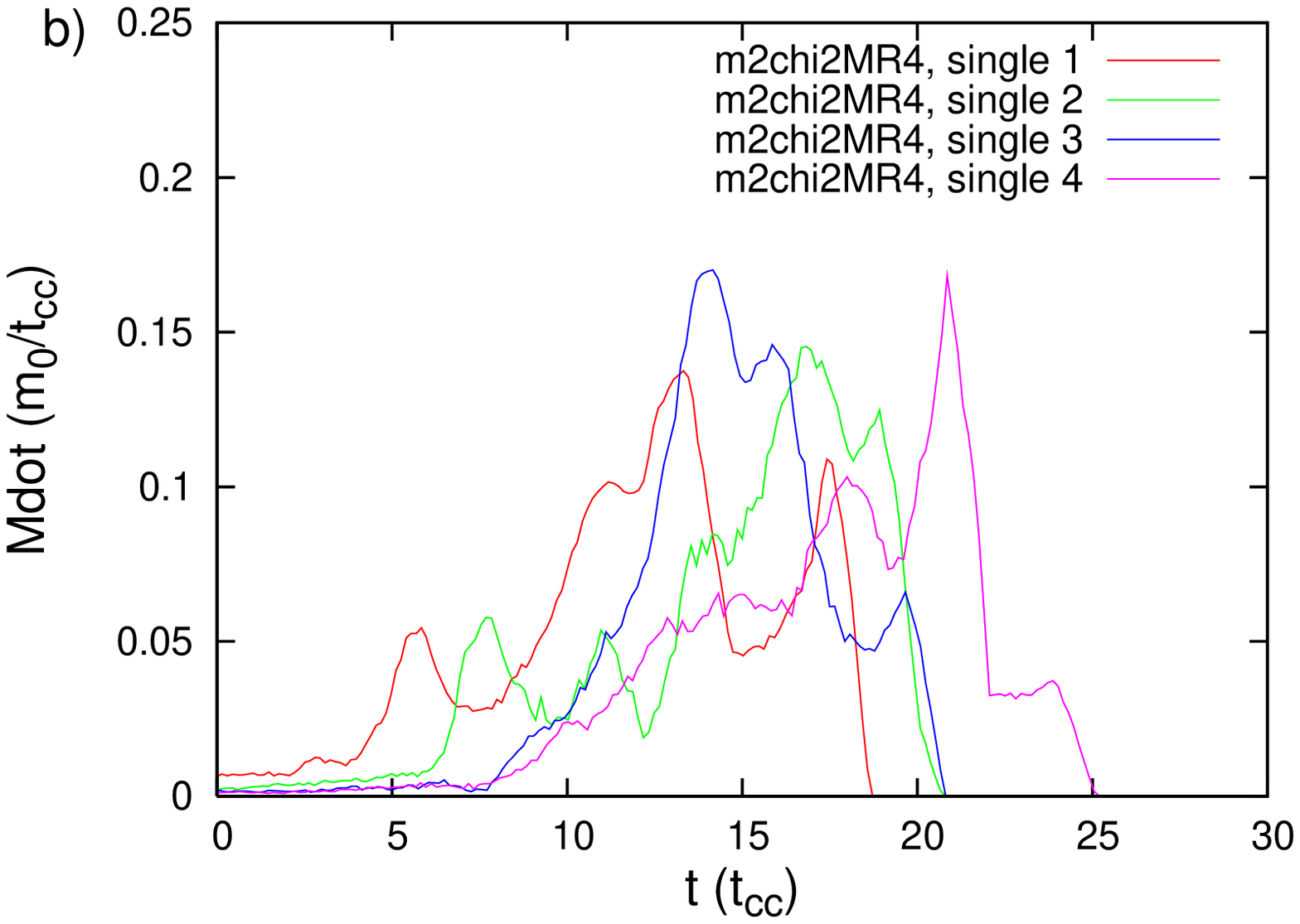, angle=-90, width=6.5cm}
\psfig{figure=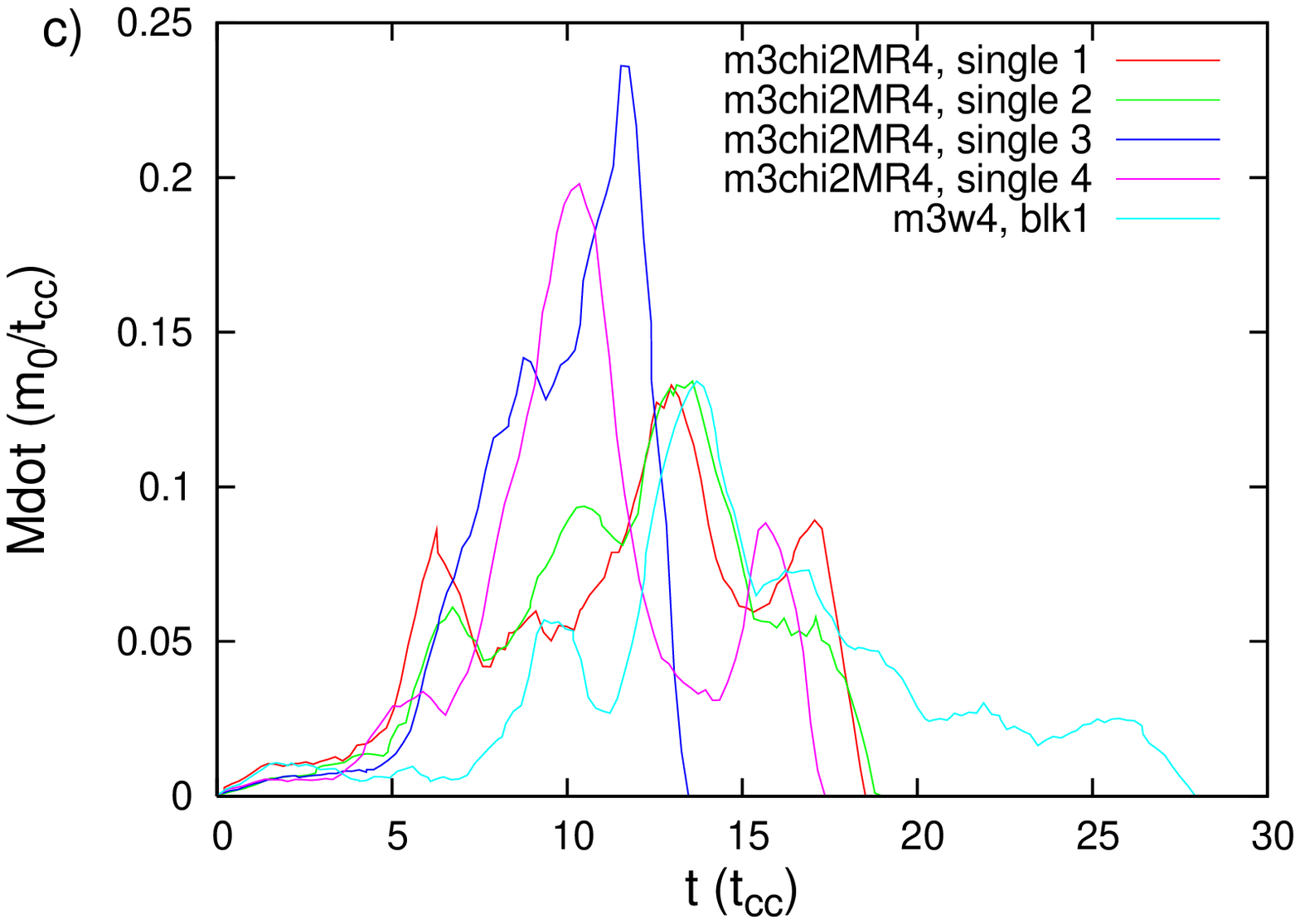, angle=-90, width=6.5cm}
\psfig{figure=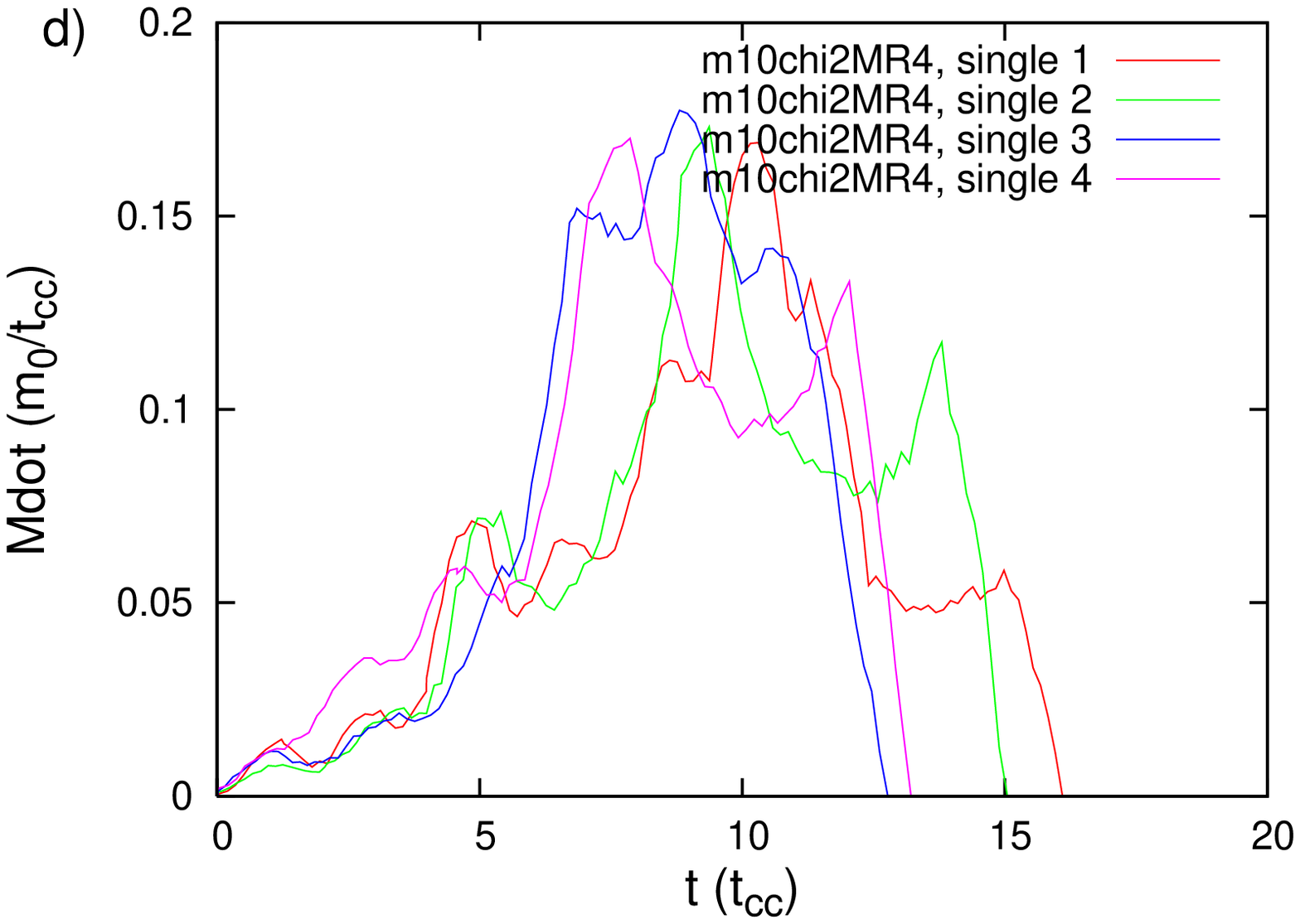, angle=-90, width=6.5cm}
\caption[]{Normalized mass loss rates for select individual clouds in
  simulations \emph{chi2MR4} for different Mach numbers: a) $M=1.5$, b)
 $M=2$, c) $M=3$ and d) $M=10$.
}
\label{fig:wide6_mdot}
\end{center}
\end{figure*}

Mass loss rates from different clouds in the simulations are shown in
Fig.~\ref{fig:wide6_mdot}. Each cloud is from a particular block (see,
e.g., Fig.~\ref{fig:m3w6_early}), and the time in each plot has been
shifted so that $t=0$ corresponds to the time when the
shock first encounters the cloud under consideration. The greatest
differences between the evolution of different clouds in a given
simulation occur in model {\em m1.5chi2MR4}, as shown in
Fig.~\ref{fig:wide6_mdot}a). Here the cloud in the block which is
first to be hit by the shock (labelled single1) encounters a flow
which is relatively unmodified by the small number of clouds which lie
further upstream. Although the interaction is relatively weak because
of the modest shock Mach number, significant mass-loss from the cloud
occurs after $t=\,5 t_{\rm cc}$, and the core of the cloud is
completely destroyed by $t\approx27\,t_{\rm cc}$ \citep[a similar
lifetime occurs for spherical clouds - see][]{2010MNRAS.405..821P}. In
contrast, significant mass-loss does not occur until
$t\approx10\,t_{\rm cc}$ for the cloud in the second block (labelled
single2), and until $t\approx12\,t_{\rm cc}$ and $13\,t_{\rm cc}$ for
clouds in the third and fourth blocks (labelled single3 and single4,
respectively). The low initial mass-loss rates from clouds further
downstream is due to the decay of the shock into a wave as it moves
through the clumpy medium. While the flow past the cloud is subsonic
the mass-loss rate remains very low.

The same initial behaviour is also seen in simulation {\em m2chi2MR4},
as shown in Fig.~\ref{fig:wide6_mdot}b). In contrast, at higher shock
Mach numbers (models {\em m3chi2MR4} and {\em m10chi2MR4} in
Fig.~\ref{fig:wide6_mdot}) the shock remains supersonic as it transits
through the clumpy region, and differences between the clouds in the
evolution of their mass-loss rates are not so readily apparent (see
Fig.~\ref{fig:wide6_mdot}c and d).

\begin{figure}
\begin{center}
\psfig{figure=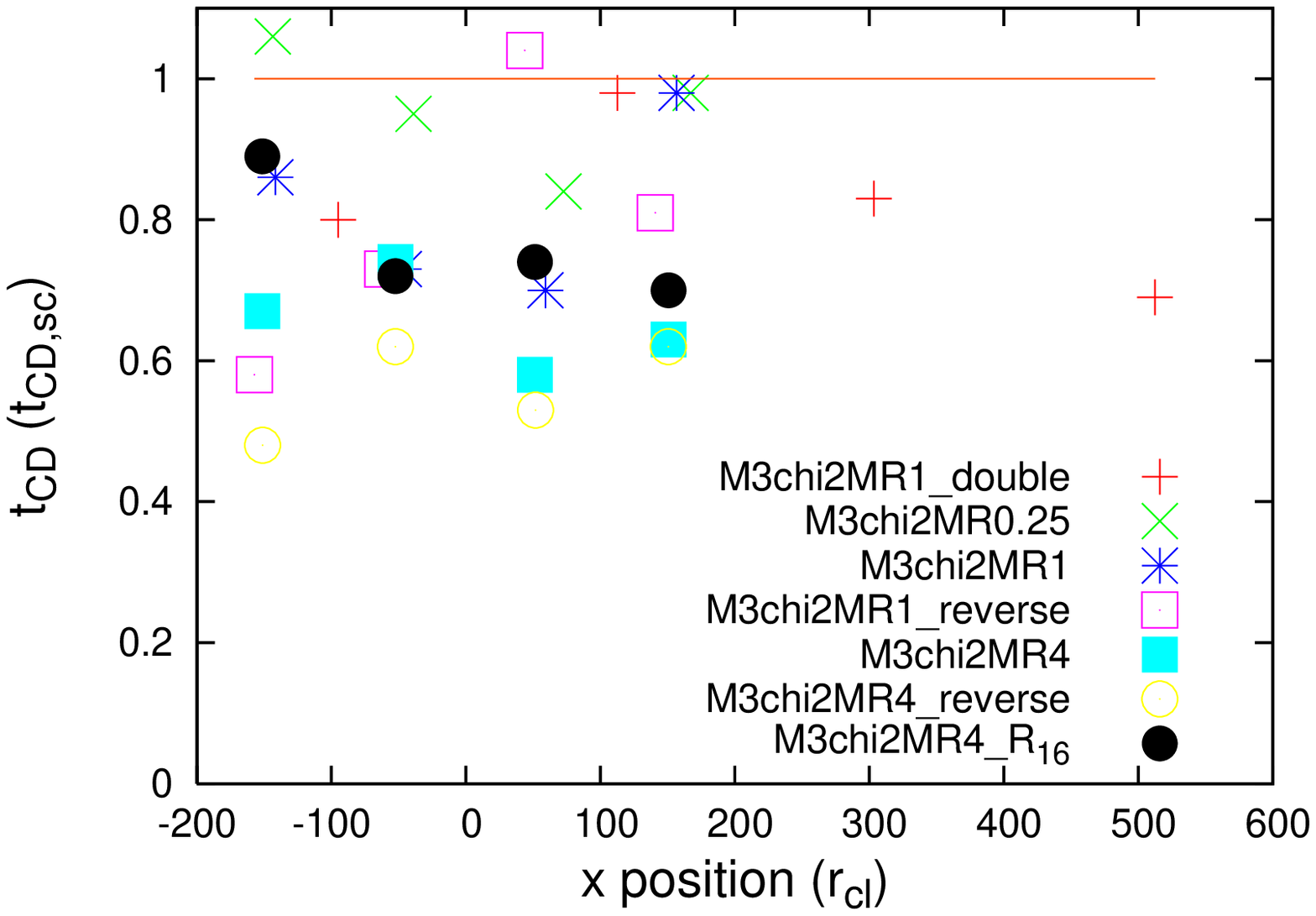, angle=-90, width=6.5cm}
\psfig{figure=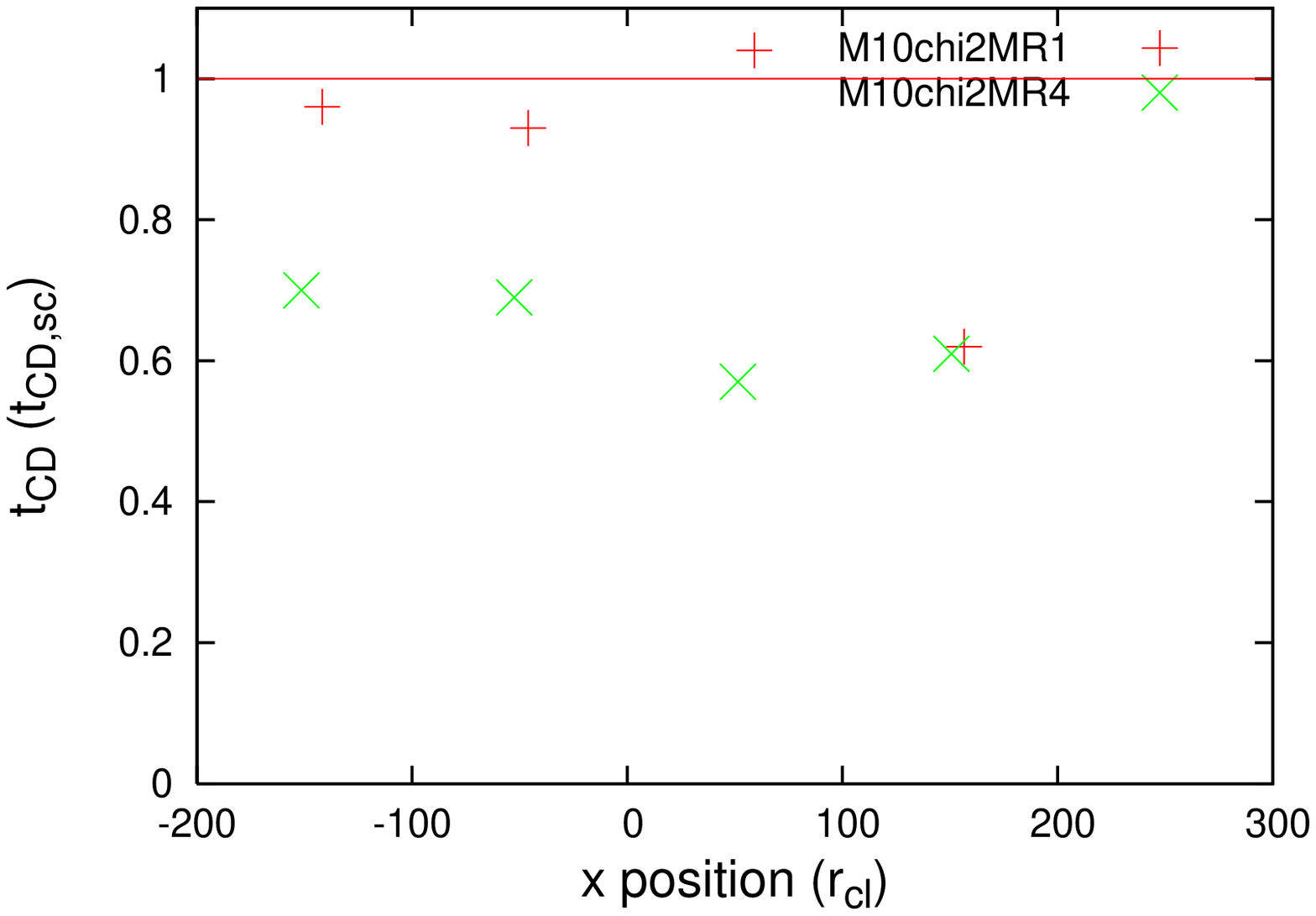, angle=-90, width=6.5cm}
\caption[]{Ratio of the lifetime of a downstream cloud to the
  equivalent lifetime obtained when the cloud is hit by a ``clean''
  shock with a uniform post-shock flow.
  }
\label{fig:statistical_lifetimes}
\end{center}
\end{figure}

It is clear that the cloud lifetimes must be compared in a statistical
way. Fig.~\ref{fig:statistical_lifetimes} shows the ratio of the
lifetimes of individual clouds to the lifetime of the equivalent cloud
hit by a ``clean'' shock (i.e. initially with no post-shock structure
due to interactions with upstream clouds), $R_{\rm life}$. The top
panel in Fig.~\ref{fig:statistical_lifetimes} shows the results for
$M=3$, while the bottom panel shows the results for $M=10$. In both
cases there is a general trend for a reduction in $R_{\rm life}$ with
increasing $MR$, albeit with a reasonably large scatter for a
particular cloud. Table~\ref{tab:statistical_lifetimes} notes the
average value of $R_{\rm life}$ in a given simulation, where this
trend is now even clearer. When $MR=4$, we find that downstream clouds
have lives which are about 40\% shorter than the lifetime of an
isolated.  Table~\ref{tab:statistical_lifetimes} also indicates that
there is no clear trend of $R_{\rm life}$ with the shock Mach number
$M$ for a given value of the mass-ratio, $MR$.  Finally, a simulation
performed at twice our standard resolution reveals that there is
currently a slight resolution dependence on $R_{\rm life}$ (model {\em
  m3chi2MR4} returns $R_{\rm life} = 0.62\pm{0.07}$ and
$0.72\pm{0.02}$ at a resolution of 8 and 16 cells per cloud radius,
respectively).

The reduction in the cloud lifetime due to the turbulent nature of the
flow is clearly more significant than the weakening and broadening of
the shock which acts to increase the cloud lifetime \citep{2002MNRAS.333....1W}. 

\begin{table}
\caption{Ratio of the lifetimes of a downstream cloud to the
  equivalent lifetime of an ``isolated'' cloud. The quoted ratio is
  averaged over the lifetime of the individual cloud tracked in each
  block of clouds.} 
\centering 
\begin{tabular}{c c} 
\hline\hline 
Simulation & $R_{\rm life}$ \\ [0.5ex] 
\hline 
m2chi2MR4 & $0.55\pm{0.07}$ \\
m3chi2MR0.25 & $0.92\pm{0.07}$ \\
m3chi2MR1 & $0.83\pm{0.14}$ \\
m3chi2MR4 & $0.62\pm{0.07}$ \\
m3chi2MR4\_R16 & $0.72\pm{0.02}$ \\
m10chi2MR1 & $0.86\pm{0.22}$ \\
m10chi2MR4 & $0.62\pm{0.06}$ \\[1ex] 
\hline 
\end{tabular}
\label{tab:statistical_lifetimes}
\end{table}

\section{Discussion}

Our work is relevant to objects where hot, diffuse gas interacts with a colder dense phase.  
In many of these objects the cold phase, despite its low volume filling fraction, may dominate 
the dynamics of the hot phase, and thus significantly change the morphology and evolution of 
the object, and its emission.  On the smallest scales these objects include wind-blown bubbles
 and supernova remnants. \cite{1981ApJ...247..908C} 
were the first to study the behaviour 
of supernova remnants expanding into a clumpy medium. They found that the destruction of the 
clouds leads to the highest densities in the remnant occuring over the outer half radius (in 
contrast, when there is no mass loading, a thin dense shell forms at the forward shock). 
These findings have since been supported by \cite{1987MNRAS.228..453D}, 
who reported a similar "thick shell" morphology in their similarity solutions, and by 
the additional numerical simulations presented by \cite{2002A&A...390.1063D} 
and \cite{2003A&A...401.1027P}. 
The X-ray emission in these cases becomes 
softer and more extended.  In other work, \cite{1996ApJ...457..752A} 
studied the effects 
of mass loading by hydrodynamic ablation on supernova remnants evolving
inside cavities evacuated by the stellar winds of the progenitor
stars. They showed that the extra mass injected by embedded clumps was
capable of producing the excess soft X-ray emission seen in some
bubbles in the Large Magellanic Cloud. We conclude, therefore, that cloud destruction 
by ablation, as in the simulations presented in our paper, can be looked for by searching 
for its affect on the X-ray emission of supernova remnants.

The interaction between a wind and individual clouds is also seen in observations of
planetary nebula. For instance, in
NGC~7293 (the Helix nebula)
long molecular tails and bright cresent-rimmed clouds are
spectacularly resolved \citep{2005AJ....130..172O,2006ApJ...652..426H,2007MNRAS.382.1447M}. 
The tails in the outer part of the nebula are less clear due to projection
effects, but appear to be a separate population displaying wider
opening angles. This may reflect changes in the diffuse flow past the
clouds, possibly due to the material stripped off the clouds further
upstream. Observations indicate that the flow past the clouds
is mildly supersonic. Although numerical simulations are able to match the
basic morphology of the tails \citep{2006A&A...457..561D}, dedicated 3D simulations
of clouds with very high density contrasts to the ambient flow are
required in order for further insight to be gained.

On larger scales, we note that there is substantial support
for mass-loading in starburst regions. Broad emission-line
wings are seen in many young star forming regions including 30 Doradus
\citep{1994ApJ...425..720C, 1999MNRAS.302..677M}, NGC\,604
\citep{1996AJ....112..146Y}, and NGC\,2363 \citep{1992ApJ...386..498R,
  1994ApJ...437..239G}, and more distant dwarf galaxies
\citep[e.g.][]{1995ApJ...438..563M, 1996ApJ...458..524I,
  1999ApJ...522..199H, 2006MNRAS.370..799S}. More recently,
\citet{2007MNRAS.381..894W, 2007MNRAS.381..913W} reported on the broad
emission-line component in the dwarf irregular starburst galaxy
NGC\,1569. Although the nature of the broad lines has yet to be fully
determined, evidence is mounting that it is associated with the impact
of cluster winds on cool gas knots. It therefore traces both
mass-loading of wind material and mass entrainment.
Further support for mass-loading comes from the analysis of hard X-ray
line emission within starburst regions. \citet{2009ApJ...697.2030S} determine that as much
material is mass-loaded into the central starburst region of M82 as is
expelled by the winds and supernovae which pressurize the region.
A key future goal is the development of numerical simulations of the multi-phase gas 
within starburst regions, and predictions for broad emission-line wings and X-ray emission.

Starbursts are often associated with galactic outflows. These flows are observed to be filamentary.
 The consensus view is that the clumps at the heads of the filaments are material which is ripped 
out of the galactic disk as the galactic wind develops (i.e. representing additional, distributed,
 mass-loading). In some cases the filaments appear to be confined to the edges of the outflow 
\citep{1998ApJ...493..129S}, while in other objects they appear to fill the interior of the wind 
\citep{1997ApJ...479L.105V}. It is clear that material stripped from the H$\alpha$ emitting clouds 
is entrained into the outflow \citep[see, e.g.][]{2008ApJ...674..157C}, but the exact amount is notoriously 
difficult to measure \citep{2005ARA&A..43..769V}. Again, future dedicated simulations are needed to help address this issue.

A key question concerns the ultimate fate of gas within
galactic outflows. Observations and simulations indicate that the majority of
the energy in galactic winds is in the kinetic energy of the hot gas,
while the mass in the outflow is dominated by the warm photoionized
gas. The former has a good chance of escaping the gravitational
potential of the host galaxy, while the latter in many cases is
unlikely to do so. Even when simple energetic arguments suggest that
the ISM can be completely expelled from a starbursting galaxy, whether
it will actually occur depends strongly on the geometry and multiphase
nature of the ISM \citep[see, e.g.,][]{1995ApJ...448...98H}. For
instance, if a centrally concentrated starburst occurs in a galaxy
with a disk-like ISM, blowout of the superbubble along the minor axis
can allow the bulk of the ISM in the disk to be retained by the
galaxy. With a multiphase ISM the diffuse intercloud medium may be
ejected while large dense clouds remain in the disk.

Recent work by \cite{2008MNRAS.387..577O} 
indicates that the range of
galactic winds is primarily determined by the interaction of the wind
with the ambient environment, with the gravity of the parent galaxy
playing a less significant role. Their simulations also show that across cosmic
time the average wind particle has participated in a wind several
times. However, further investigations are needed, since their simulations currently lack the resolution
required to make accurate quantitative predictions of the slowing of the winds and wind recycling.
Furthermore, the wind's
multiphase nature must be addressed. Studies like ours are relevant to this work
since the destruction and
acceleration timescales which we find for our clumpy region have some bearing on
the mixing and stalling timescales of a galactic wind. Such simulations could be tested 
against observations of the extent of galactic winds 
\citep[e.g.][]{2011Sci...334..948T, 2011Sci...334..952T}.

\section{Conclusions}
\label{sec:con}
We have performed a detailed investigation of a shock running
through a clumpy region. We find the following key behaviour:

\begin{itemize}
\item The stripping of material from the clouds ``mass-loads'' the
  post-shock flow and leads to the formation of a dense
  shell. Fully-mixed material within the shell reaches a maximum
  density, after which the shell grows in width. The shell expands and
  its density drops once there are no more clouds to mix in.

\item The evolution of the shock can be split into several distinct
  stages. During the first stage, the shock decelerates. Then in some
cases its velocity becomes nearly constant. After deceleration and the constant 
speed phase, if it occurs, the shock accelerates and finally approaches
the speed it had in the uniform medium before it encountered the clumpy region.
 A steady stage does not always occur (e.g., if the clumpy region is not very deep).

\item When the mass-loading is sufficient, the flow can be 
slowed to the point that the shock degenerates into a wave.

\item The clumpy region becomes more porous as the number density of
  clouds is reduced, which occurs for lower values of
  $MR$, and/or increased values of $\chi$.

\item A great deal of turbulence is generated in the post-shock flow
  as it sweeps through the clumpy region. Clouds exposed to this
  turbulence can be destroyed in only 60\% of the time needed to
  destroy a similar cloud in an ``isolated'' environment. The lifetime
  of downstream clouds decreases with increasing $MR$.
\end{itemize}

We have determined the necessary conditions (in terms of the cloud
density contrast and the ratio of cloud mass to intercloud mass) for a
clumpy region to have a significant effect on a diffuse flow. The
lifetime of clouds is a key factor in this respect.
\citet{2009MNRAS.394.1351P} first showed that clouds can be destroyed more
quickly when overrun by a highly turbulent environment, although the
strength of this turbulence was treated as a free parameter. In
contrast, in this paper we have presented the first self-consistent
simulations of a highly turbulent flow overrunning clouds, where the
enhanced turbulence is a natural consequence of the flow overruning
clouds further upstream. Since our simulations are purely
hydrodynamic, the next step will be to investigate the effects of
magnetic fields and thermal conduction on the flow dynamics.
In future papers we will also investigate the ability of a flow to force its
way through a finite-sized clumpy medium, and determine how this 
depends on the ratio of the mass injection rate from the clouds to the
mass flux in the wind.

\section*{acknowledgements}
JMP would like to thank the Royal Society for funding a
University Research Fellowship.
We would also like to thank the referee for producing a constructive and timely report.


\bibliography{biblio}

\end{document}